\documentclass[letterpaper,twocolumn,british,english]{IEEEtran}
\usepackage{array}
\usepackage{float}
\usepackage{mathrsfs}
\usepackage{amsthm}
\usepackage{amsmath}
\usepackage{amssymb}
\usepackage{graphicx}
\usepackage[numbers]{natbib}

\makeatletter

\pdfpageheight\paperheight
\pdfpagewidth\paperwidth

\providecommand{\tabularnewline}{\\}
\floatstyle{ruled}
\newfloat{algorithm}{tbp}{loa}
\providecommand{\algorithmname}{Algorithm}
\floatname{algorithm}{\protect\algorithmname}

\numberwithin{equation}{section}
\numberwithin{figure}{section}
\theoremstyle{plain}
\newtheorem{thm}{\protect\theoremname}
\theoremstyle{definition}
\newtheorem{defn}[thm]{\protect\definitionname}
\newenvironment{lyxlist}[1]
{\begin{list}{}
{\settowidth{\labelwidth}{#1}
 \setlength{\leftmargin}{\labelwidth}
 \addtolength{\leftmargin}{\labelsep}
 }}
{\end{list}}

\usepackage{subfig}
\usepackage{url}
\usepackage{algorithm}
\usepackage{algpseudocode}
\newcommand{\paper}[1]{paper#1}

\makeatother

\usepackage{babel}
\addto\captionsbritish{\renewcommand{\algorithmname}{Algorithm}}
\addto\captionsbritish{\renewcommand{\definitionname}{Definition}}
\addto\captionsbritish{\renewcommand{\theoremname}{Theorem}}
\addto\captionsenglish{\renewcommand{\definitionname}{Definition}}
\addto\captionsenglish{\renewcommand{\theoremname}{Theorem}}
\providecommand{\definitionname}{Definition}
\providecommand{\theoremname}{Theorem}

\begin{document}

\title{Measuring Privacy Leakage\index{measuring privacy leakage} for IDS
Rules}

\author{\IEEEauthorblockN{Nils Ulltveit-Moe and Vladimir Oleshchuk}\\
\IEEEauthorblockA{Department of Information and Communication Technology\\
University of Agder\\
Jon Lilletuns vei 9, 4879 Grimstad, Norway\\
Email: \{Nils.Ulltveit-Moe, Vladimir.Oleshchuk\}@uia.no}}
\maketitle
\begin{abstract}
\emph{}This \paper{}  proposes a measurement approach\index{measurement approach}
for estimating the privacy leakage\index{privacy leakage} from Intrusion
Detection System (IDS) alarms. Quantitative information flow analysis\index{quantitative information flow}
is used to build a theoretical model of privacy leakage from IDS rules,
based on information entropy. This theoretical model is subsequently
verified empirically both based on simulations and in an experimental
study. The analysis shows that the metric is able to distinguish between
IDS rules that have no or low expected privacy leakage and IDS rules
with a significant risk of leaking sensitive information, for example
on user behaviour. The analysis is based on measurements of number
of IDS alarms, data length and data entropy for relevant parts of
IDS rules (for example payload). This is a promising approach that
opens up for privacy benchmarking\index{privacy benchmarking} of
Managed Security Service providers.
\end{abstract}

\section*{Keywords}

Intrusion detection, privacy leakage, entropy, metrics

\section{Introduction}

\emph{}

The objective of this \paper{}  is to develop an entropy-based metric
that can be used for privacy leakage detection \index{model of privacy leakage}
in intrusion detection system (IDS) alarms. The approach should be
able to identify IDS rules\index{IDS rules} that according to stakeholders'
perception have a significant potential for leaking private or confidential
information. It should also identify the worst IDS rules from a privacy\index{privacy}
or confidentiality\index{confidentiality} perspective based on indicators
that can be calculated automatically. For example IDS rules that:
\begin{itemize}
\item have a significant risk to leak information that is sensitive (privacy
sensitive\index{privacy sensitive}, security sensitive\index{security sensitive},
business sensitive\index{business sensitive} etc.);
\item have an unclear or too simple definition of the attack detecting pattern\index{attack detecting pattern},
and therefore may trigger unnecessarily, in the worst case on person
sensitive or confidential information.
\end{itemize}

Privacy policies can be used to define what information that is sensitive\index{sensitive}.
Examples of sensitive information may be certain IP ranges of classified
systems or sampled payload that may reveal private or confidential
information. Information can also be defined as person sensitive by
law, for example the sampled payload from a health institution which
may contain person sensitive information. Another example is critical
infrastructures that may contain security sensitive or confidential
information in the data traffic about the processes being controlled.
Last, but not least, payment databases handling financial transactions
may reveal sensitive information like credit card numbers.

In these cases, the information is \emph{per definition} sensitive,
which means that \emph{any} leakage of information that can be identified
may be problematic. For such use cases, an objective information leakage
metric will be sufficient to identify problematic leakage of private
or confidential information\index{confidential information}.

In other cases, the privacy sensitivity will be subjective, and can
only be evaluated in a representative way by the\emph{ owners} of
the data being sampled - the users themselves. It may even in this
case be possible for the data controller\index{data controller} to
get realistic estimates of the perceived privacy sensitivity by asking
a representative random set of users, for example using a random poll
on the service being used, about how they would value privacy leakages.
However this approach will be expensive and does not scale well. It
is therefore only viable for smaller evaluations of privacy impact.

It is therefore assumed possible for an authority like the data controller,
that is overseeing the privacy interests, to estimate the privacy
impact\index{privacy impact}, denoted by $I\geq0$, that an identified
information leakage $L\geq0$ causes. The privacy impact could for
example be the subjective value or expected liability from privacy
or confidentiality breaches, as proposed by \citep{gritzalis_probabilistic_2007}.
The privacy leakage, denoted by $\pi_{R}$ for a given IDS rule $R$
can then be defined as the product of the information leakage metric
$L$ and the privacy impact $I$, i.e: $\pi_{R}=I\cdot L$. However,
if investigation shows that the information leakage is caused by activities
from attack vectors that do not cause any risk of revealing private,
business sensitive or confidential information, then the privacy impact
for a given IDS rule may be set low or even to zero. The combined
metric $\pi_{R}$ can be regarded as a privacy leakage risk metric\index{risk metric},
that can be used to measure and perform incremental improvements of
the Managed Security Service (MSS) operation from a privacy perspective.

Current IDSs typically provide an all or nothing solution for handling
private or confidential information in the alarms. The payload of
the alarms is either being sent in cleartext or may be pseudonymised,
for example by only sending references to where more information can
be found in a data forensics system. There does not exist a more fine-grained
management nor any measurements of sensitive information flows in
such systems. It is in particular common that Open Source based IDS's
like Snort, OSSEC or Prelude send payload in cleartext in the IDS
alarms. Having a metric for how privacy invasive\index{privacy invasive}
an MSS operation is will therefore be useful to benchmark the performance
of different MSS providers from a privacy perspective. It will also
be useful for tuning the IDS rulesets and for implementing anonymisation
policies to reduce the privacy impact of the monitoring. Intuitively,
such a privacy leakage model relates to the perceived \emph{preciseness\index{preciseness}}
of the IDS rule, i.e. how good it is at detecting only attack traffic
without revealing non-attack traffic. 

A promising candidate for a privacy leakage metric for IDS rules,
is data entropy\index{entropy}. This is a privacy leakage metric
that is based on the \emph{variability} of the underlying data. Examples
of such metrics are Shannon-, Rényi or Min-entropy, which previously
have been proposed as anonymity metrics~\citep{shannon_1948,clau_structuring_2006}.
Entropy can also be used to measure \emph{coding efficiency\index{coding efficiency}},
for example whether sampled payload excerpts most likely are encrypted
or compressed~\citep{shannon_1948}. This \paper{}  investigates
a model of privacy leakage from IDS rules that is based on the variation
in entropy\index{variation in entropy} between IDS alarms. This is
to the best of our knowledge the first comprehensive privacy leakage
model for IDS rules based on quantitative measurements of information
flow founded in information theory\index{information theory}.

The proposed privacy leakage metric has several practical applications\index{practical applications}.
First, it can be used to identify imprecise IDS rules, since such
rules typically will have more variation in the underlying data, and
therefore also a larger variance in entropy than more precise IDS
rules. Furthermore, an advantage with the proposed metric is that
it can detect two common ways of preserving privacy or data confidentiality:
anonymisation and pseudonymisation. Both encrypted and anonymised
information can be expected to have zero entropy variance, given sufficiently
long input. On the other hand, the entropy variance of plaintext data
will be significantly larger than for encrypted data, as will be discussed
in Section \ref{sub:Difference-Encrypted-Plaintext}. 

This means that the entropy variance can be used as a metric to detect
leakage of private or confidential information\index{detect leakage of private or confidential information}
in message oriented data streams in general and IDS alarms in particular.
It can also be used to verify whether an anonymisation/pseudonymisation
or encryption scheme works as intended. 

This \paper{}  is organised as follows: Section \ref{sec:Motivation}
discusses the motivation behind introducing an entropy variance based
information leakage metric, based on existing knowledge of how common
attack vectors work. Section \ref{sec:Threat-Model} describes the
threat model and scenario that is assumed when using the privacy leakage
metric. Section \ref{sec:A-Privacy-Leakage-Model} develops the entropy-based
privacy leakage model based on quantitative information flow analysis
after introducing the necessary theoretical background information.
The last part discusses how clustering based on the Expectation Maximisation
algorithm can be used to identify the underlying attack vectors for
IDS rules that detect more than one attack vector. Section \ref{sec:Detailed-Analysis-of}
does a detailed analysis of the convergence speed as a function of
amount of input data for the entropy algorithms and symbol definitions
considered. This includes analysing the metrics' abilities to distinguish
between plaintext and encrypted data. Section \ref{sub:Analysis-of-Alarm-PDF}
analyses experimental results based on realistic measurements of IDS
alarms. Section \ref{sec:Related-Works} discusses related works;
Section \ref{sec:Conclusion} concludes the \paper{}  and Section
\ref{sec:Future-Work} suggests future work and research opportunities.

\section{\label{sec:Motivation}Motivation\index{motivation}}

A precise IDS rule will in many cases report only one or a few different
attack patterns corresponding to real attack vectors, as will be discussed
below. One common type of attack vector that follows this behaviour,
is stack or heap buffer overflow attacks~\citep{vallentin_evolution_2007}.
These attack vectors\index{attack vectors} frequently use large sequences
of characters corresponding to the NOP operation\index{NOP operation}
or similar to increase the probability of successfully exploiting
buffer overflow vulnerabilities\index{exploit buffer overflows}.
The attacker does then not need to know the exact memory location
of injected shellcode\index{shellcode}, since returning to any address
within the NOP sled\index{NOP sled} will cause the shellcode to be
executed. This makes it simpler for the adversary to exploit such
vulnerabilities. The entropy of this NOP sled will be zero, and variance
zero, as long as only NOP operations are being used in the sled and
the attack vector does not mutate (e.g. by changing the length of
the NOP sled). This is clearly distinguishable from ordinary traffic,
and also easy to distinguish for rule-based IDSs.

Such naive attacks are however not so common nowadays, because the
IDS and anti-virus\index{anti-virus} technologies easily can detect
such anomalies in the input. It is therefore increasingly common that
the adversaries obfuscate the attack vector. Obfuscation\index{obfuscation}
of the NOP sled can for example be done using metamorphic coding\index{metamorphic coding},
which means that instructions in the sled are substituted with other
instructions that effectively perform the same function~\citep{jordan_writing_2005}.
Furthermore, it is now common practice that also the shellcode\index{shellcode}
of the attack is being obfuscated by using encryption techniques.
This means that the attack after the NOP sled contains a small decryption
program, with a decryption key that decrypts the obfuscated shellcode
before it is being run~\citep{song_infeasibility_2007}. Even the
decryption program can be hidden by using metamorphic coding techniques~\citep{song_infeasibility_2007},
although this is still not very common~\citep{polychronakis_empirical_2009}.

This means that obfuscated attack vectors can be expected to have
\emph{quite high entropy}, in some cases \emph{indistinguishable}
from encrypted traffic~\citep{song_infeasibility_2007,goubault-larrecq_detecting_2006}.
This means that the \emph{variation in entropy\index{variation in entropy}}
can be expected to go towards zero for a sufficiently large data sample
from a polymorphic attack vector, given that it is indistinguishable
from a perfect encryption scheme. Such an attack vector\index{attack vector}
will behave like random uniform data.
This means that the entropy variance of sufficiently large attack
vector samples from both traditional NOP sled based attacks and modern
obfuscated attacks also can be expected to have \emph{low} entropy
variance. 

It can furthermore be observed that samples of encrypted user traffic\index{encrypted user traffic},
assuming that strong encryption is used, in itself does not leak any
private or confidential information, hence can be expected to have
low entropy variance\index{entropy variance}. Ordinary non-encrypted
user traffic\index{non-encrypted user traffic}, can however be expected
to show a significant variance in entropy between different samples,
as illustrated in Figure \ref{fig:Difference-bit-entropy}. This indicates
that entropy variance may be an interesting metric for measuring whether
IDS alarms leak information, in particular for buffer overflow\index{buffer overflow}
type of attacks. However this metric does obviously not understand
the semantics \index{semantics}of the data traffic, and can therefore
not be used to evaluate whether the leaked information is private
or confidential.

There also exist attack vectors that are indistinguishable from plaintext
data. Examples of such attacks are nonobfuscated Javascript Trojans\index{Trojans}
or SQL injection attacks\index{SQL injection attacks}. This means
that the entropy standard deviation not necessarily can be assumed
to be close to the extreme points: encrypted data (entropy close to
1) or NOP sleds (octet-entropy close to 0). However, there are still
some other useful characteristics of such plaintext attacks in particular,
and malware in general, that can be exploited by such a metric:
\begin{itemize}
\item Attacks are to a great extent automated and performed by large botnets\index{botnets}
of compromised hosts.
\item Attack vectors do typically not yet mutate or change dynamically%
\footnote{Although proof-of-concept polymorphic self-mutating worms has been
demonstrated~\citep{kolesnikov_advanced_2005}.%
}. This means that multiple attacks by a given host being controlled
by an adversary typically has the same payload. Different hosts running
the same version of a given malware\index{malware} can also be expected
to typically have the same payload~\citep{polychronakis_empirical_2009}.
\item Attack vectors are modular programs that are improved incrementally,
which means that not all parts of a malware will change at the same
time, and some parts of malware code are even shared between different
malware families~\citep{polychronakis_empirical_2009}.
\item Botherders\index{botherders}, that manage large botnets of compromised
hosts, will also have a self interest in a ``well managed'' botnet.
This means that the malware of a botnet at regular intervals will
be upgraded to include patches and new functionalities, amongst others
to avoid being detected by Anti-Virus and IDS~\citep{freiling_computer_2005}.
It is therefore reasonable to believe that a large amount of the machines
in a given botnet will run the same version of the malware and therefore
also will use the same arsenal of attack vectors\index{attack vectors}
for attacking other hosts.
\end{itemize}
This means that if an IDS rule is able to detect a given attack, or
attack variants, then there are several reasons to believe that the
entropy variance between instances of the same attack vector may be
small, even for nonobfuscated Javascript or SQL injection attacks.
This furthermore means that if the underlying attack vectors detected
by an IDS rule can be identified, then the entropy variance (or entropy
standard deviation) around each attack vector can be considered a
measure of the precision\index{precision} of that rule hence also
an indicator of possible privacy leakage\index{privacy leakage}s.

\section{\label{sec:Threat-Model}Threat Model\index{threat model}}

\emph{}

The \paper{}  assumes that intrusion detection services have been
outsourced\index{outsourced} to a third party Managed Security Service
(MSS) provider. Security monitoring is furthermore subdivided into
two different security levels. An outsourced first-line service\index{first-line service}
that is doing 24x7 monitoring\index{24x7 monitoring} of the computer
networks, and a trusted second-line service\index{trusted second-line service}
that will have full knowledge of the IDS service, including capabilities
to perform data forensic analysis. It is assumed that the MSS provider
operates using a privacy-enhanced IDS, so that changes to the IDS
ruleset must be agreed upon by both the data controller\index{data controller}
and the second line security analyst\index{security analyst} responsible
for updating the IDS ruleset, to avoid that excessively privacy violating
IDS rules are being deployed. 

It is therefore assumed that the IDS services run in a controlled
environment\index{controlled environment}, where enforcement of a
privacy policy supported by privacy leakage metrics makes sense. An
example of such an environment is critical infrastructures\index{critical infrastructure}
or hospitals\index{hospitals} where security services have been outsourced
to a third party, and privacy metrics are required to ensure compliance
both to privacy and security policies. These policies must ensure
that the first-line security analysts, that are not trusted to see
sensitive information, do not get access to information considered
private or confidential by the owner of the critical infrastructure.
The objective is a stricter enforcement of the need-to-know principle\index{need-to-know principle}
than what IDSs typically have today. However, in order to enforce
such privacy and security policies, suitable privacy metrics\index{privacy metrics}
are needed, which will be developed here.

This \paper{}  mainly focuses on two adversaries\index{adversaries}:
an external adversary\index{external adversary} that may want to
manipulate the privacy metrics for example to reduce the chance of
attacks being detected. The IDS ruleset is assumed public, so that
an external adversary\index{external adversary} can investigate how
the IDS rules work in order to perform targeted attacks on either
privacy or security. However the external adversary will not know
which IDS rules that are enabled.

Insiders are divided into two main groups. First-line security analysts
are considered untrusted insiders\index{untrusted insiders}, that
only have limited authorisation to see information and no authorisation
to modify information related to the IDS configuration. They do not
have access to the data forensic\index{forensic} tool to investigate
attacks in detail. Second-line analysts are considered a trusted CERT
team\index{trusted CERT team}, that has authorisation to perform
security investigations and reconfigure the IDS. A third actor is
the data controller, who shares the responsibility for managing the
IDS ruleset with the security officer\index{security officer}, to
ensure that both the privacy and security objectives are being considered.
The \paper{}  furthermore assumes that suitable enforcement mechanisms
exist, for example anonymisation or pseudonymisation schemes for sensitive
information in IDS alarms, so that the privacy leakage metrics can
be used for verification of the security or privacy policies.

\selectlanguage{british}%

\section{\label{sec:A-Privacy-Leakage-Model}A Privacy Leakage Model\index{privacy leakage model}
of IDS Rules}

\selectlanguage{english}%
\emph{}This section will first provide an information theoretic analysis
of privacy leakage from IDS alarms, assuming a simple model of a perfect
IDS rule\index{perfect IDS rule} $R_{P}$ that does not have any
false alarms.\emph{ }This model is subsequently generalised to handle
IDS rules that may leak potentially sensitive information, and we
then show how this model corresponds to measuring the standard deviation
of entropy from the IDS rule. It is finally shown how to measure the
privacy leakage from IDS rules that detect more than one attack vector.

\subsection{Basic Definitions}

\emph{} The definitions and notation in this section give a short
introduction to quantitative information flow analysis, and is based
on~\citep{smith_foundations_2009}. It is throughout this \paper{} 
assumed that the logarithm is taken to the base 2, i.e. $log(x)$
means $log_{2}(x)$. Shannon and Min-entropy\index{Min-entropy} can
be considered instances of the more general Rényi entropy~\citep{renyi_1961},
and we therefore use the Rényi notation to describe the entropies.
Any Rényi entropy\index{Rényi entropy} metric is denoted as $H_{\alpha}(X)$,
where $\alpha$ is the entropy degree; $\alpha=1$ represents Shannon
entropy\index{Shannon entropy} and $\alpha=\infty$ represents Min-entropy.
Given an IDS rule $R$, which may leak sensitive information from
a set of input data $X$ and to a set of IDS alarms $Y$, the objective
is then to measure how much information $R$ leaks. 

Let $X$ and $Y$ be random variables whose set of possible values
are $\mathcal{X}$ and $\mathcal{Y}$ respectively. The Shannon entropy
is then defined by \citep{shannon_1948}:

\begin{equation}
H_{1}(X)={\displaystyle \sum_{x\in\mathcal{X}}P[X=x]log\frac{1}{P[X=x]}}
\end{equation}

Shannon entropy indicates the number of bits that are required to
transfer $X$ in an optimal way. The conditional entropy\index{conditional entropy}
denoted as $H_{1}(X|Y)$ indicates the expected resulting entropy
from input data $X$ given a set of IDS alarms $Y$ that pass through
the IDS rule $R$~\citep{smith_foundations_2009}:

\begin{equation}
H_{1}(X|Y)={\displaystyle \sum_{y\in\mathcal{Y}}P[Y=y]H_{1}(X|Y=y)}
\end{equation}

where

\begin{equation}
H_{1}(X|Y=y)={\displaystyle \sum_{x\in\mathcal{X}}P[X=x|Y=y]log\frac{1}{P[X=x|Y=y]}}
\end{equation}

Min-entropy is another entropy metric that is calculated based on
the worst case (maximum) symbol occurrence probability, defined as
the vulnerability\index{vulnerability} $V(X)$ that an adversary
can guess the value of $X$ correctly in one try~\citep{smith_foundations_2009}:

\begin{equation}
V(X)={\displaystyle \max_{x\in\mathcal{X}}P[X=x]}
\end{equation}

Min-entropy indicates the number of bits required to store $V(X)$,
and is defined as~\citep{smith_foundations_2009}:

\begin{equation}
H_{\infty}(X)=log\frac{1}{V(X)}
\end{equation}

The conditional min-entropy can be defined as~\citep{smith_foundations_2009}:

\begin{equation}
H_{\infty}(X|Y)=log\frac{1}{V(X|Y)}\label{eq:H_infty_X_given_Y}
\end{equation}

where

\begin{equation}
V(X|Y)={\displaystyle \sum_{y\in\mathcal{Y}}}P[Y=y]{\displaystyle \max_{x\in\mathcal{X}}P[X=x|Y=y]}
\end{equation}

It is then possible to define the information leakage\index{information leakage}
$L_{XY}$ from $X$ to $Y$ using either Shannon or Min-entropy as
proposed by~\citep{smith_foundations_2009}:

\begin{equation}
L_{XY}=H_{\alpha}(X)-H_{\alpha}(X|Y).\label{eq:information-leakage}
\end{equation}

\subsection{Perfect model IDS Rule}

\begin{figure}
\begin{centering}
\includegraphics[scale=0.45]{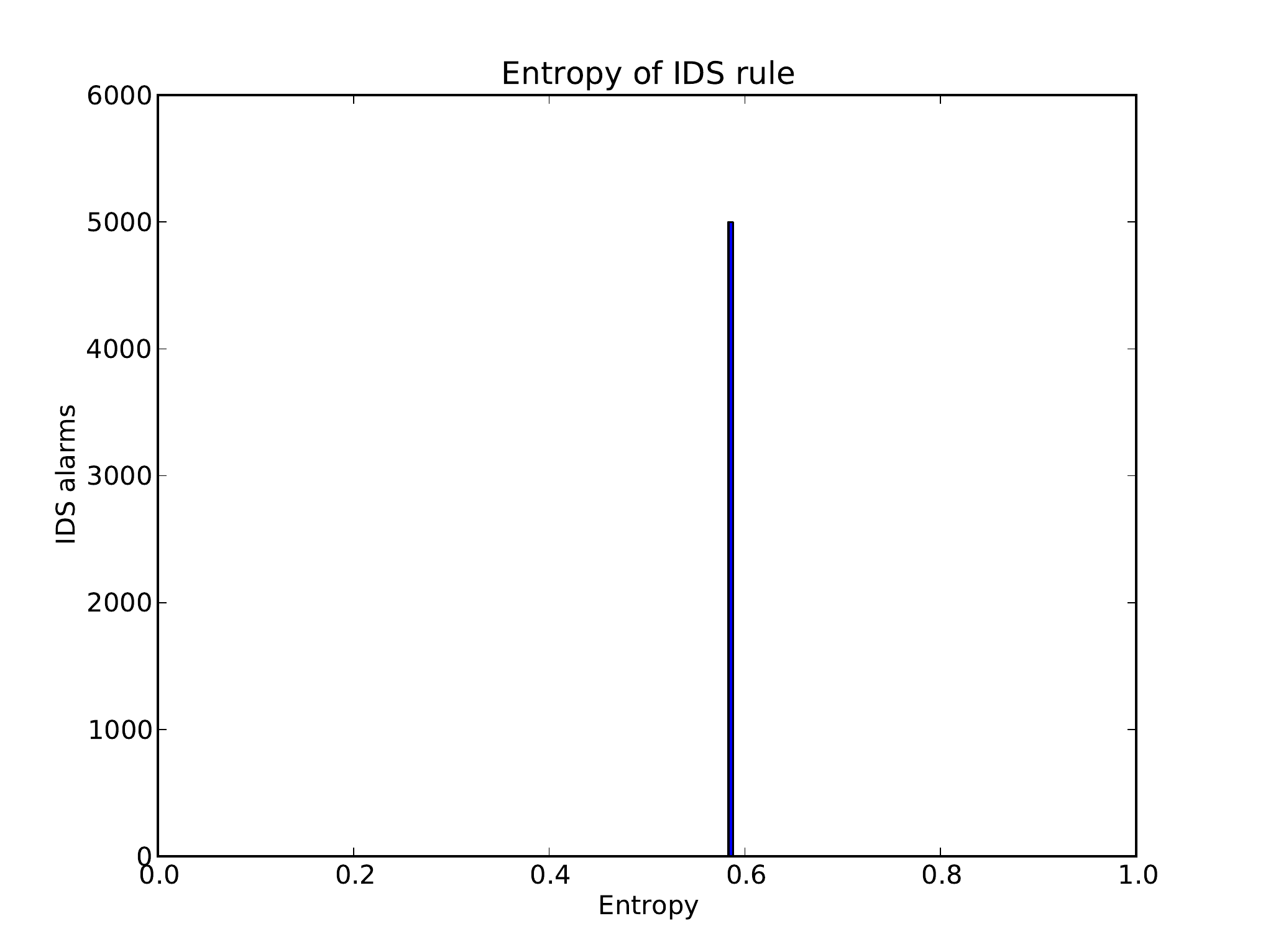}
\par\end{centering}

\caption{\label{fig:IDS-rule-1:2003}IDS rule 1:2003 SQL Worm Propagation attempt,
behaving like $R_{P}$.}
\end{figure}

Assume a perfect model IDS rule\index{perfect model IDS rule} $R_{P}$,
that always detects the attack vector and does not have any false
alarms or other entropy sources. Furthermore assume that the given
attack vector does not change between different attack instances.
The payload sample in the IDS alarm from $R_{P}$ is also assumed
to not contain any other entropy sources. The IDS will in this case
always sample the \emph{same} payload excerpt in every alarm according
to the attack pattern definition. 

This IDS rule is termed a perfect model IDS rule, since it is considered
perfect according to the theoretical model of privacy leakage. $R_{P}$
is in other words a perfect model of IDS rule \emph{behaviour}\index{rule behaviour}
from a privacy perspective. This is not a purely theoretical IDS rule
behaviour. We observed three IDS rules that behaved like $R_{P}$
in our experiments, for example the Snort IDS rule with SID 1:2003
SQL Worm Propagation attempt, as shown in Figure~\ref{fig:IDS-rule-1:2003}.
This is obviously a simplistic model of an IDS rule, since it does
not handle the fact that many IDS rules and also non-rule based technologies
like anomaly-based IDS will be able to detect more than one attack
vector, and also variants of attack vectors. The model is furthermore
oblivious to whether the source of entropies is adversarial or ordinary
user activities. An entropy-based metric can only measure whether
information is leaking or not. Therefore the privacy impact $I$ will
need to be evaluated, as discussed earlier.

The perfect model IDS rule will under these assumptions provide a
\emph{constant} leakage\emph{\index{constant leakage}} denoted as
$c$ of information in each alarm, corresponding to the pattern matched
by $R_{P}$. 

The \emph{privacy impact\index{privacy impact} $I$ }of this constant
information leakage as a \emph{privacy leakage\index{privacy leakage}}
is however not known. The privacy impact of the information leakage
from each IDS rule must therefore be evaluated by a data controller, to
determine whether the expected information leakage from the IDS rule
can be considered necessary and acceptable from a security perspective,
and also that the effective privacy impact from the rule can be considered
negligible if the rule is effective over time. 

This manual quality assurance procedure makes it possible to detect
and avoid IDS rules where $I\cdot c$ in itself is judged to cause
a significant privacy leakage, for example if the rule itself triggers
on person sensitive information. The privacy leakage $I\cdot c$ from
each installed IDS rule is therefore in the rest of this \paper{} 
considered as either necessary or negligible. If this constant privacy
leakage is not considered tolerable, then it is assumed that this
can be mitigated using anonymisation or pseudonymisation policies.

\begin{figure}
\begin{centering}
\includegraphics[scale=0.55]{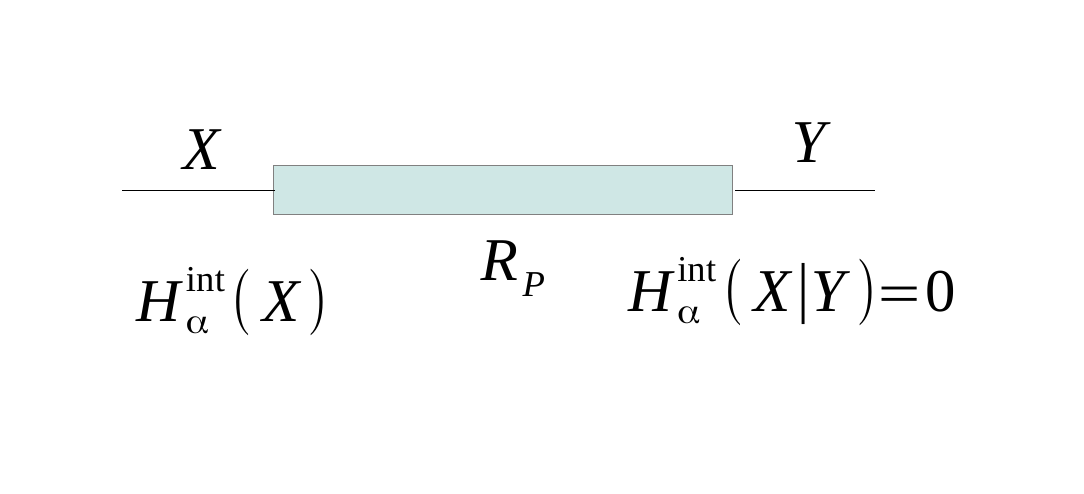}
\par\end{centering}

\caption{\label{fig:Channel-model-perfect-rule}Channel model of a perfect
model IDS rule $R_{P}$ that detects a single, nonchanging underlying
attack.}
\end{figure}

$R_{P}$ will under these assumptions always triggers on the same
attack pattern $Y=\{y\}$, as illustrated in Figure~\ref{fig:Channel-model-perfect-rule}.
The \emph{inter-alarm entropy\index{inter-alarm entropy}}, assuming
a set of input data $X$, denoted as $H_{\alpha}^{int}(X|Y)$, is
defined as the entropy between different IDS alarms, calculated over
the entire payload excerpt (i.e. each IDS alarm is considered as one
``symbol''). The inter-alarm entropy will in this case be $H_{\alpha}^{int}(X|Y)=0$,
since $P[Y=y]=1$. This means that a perfect model IDS rule according
to this definition from an \emph{information theoretical }perspective
does not reveal any \emph{additional }information apart from what
can be inferred from the limited and constant information leakage
$c$ in each alarm.

This does not mean that additional leakage of sensitive information
cannot occur, since the resulting privacy leakage also will depend
on the timing and context of the alarms. Additional information may
for example be revealed by correlating the interdependencies between
the IDS rules. 

However, under the given assumptions, this means that when $R_{P}$
triggers, then a known data pattern will have been sent in the input
data stream. This information leakage is considered a tolerable privacy
leakage under the assumptions in the previous subsection.

\subsection{\label{sub:A-Non-perfect-IDS-rule}A Non-perfect IDS rule\index{non-perfect IDS rule}
$R$}

\begin{figure}
\begin{centering}
\includegraphics[scale=0.45]{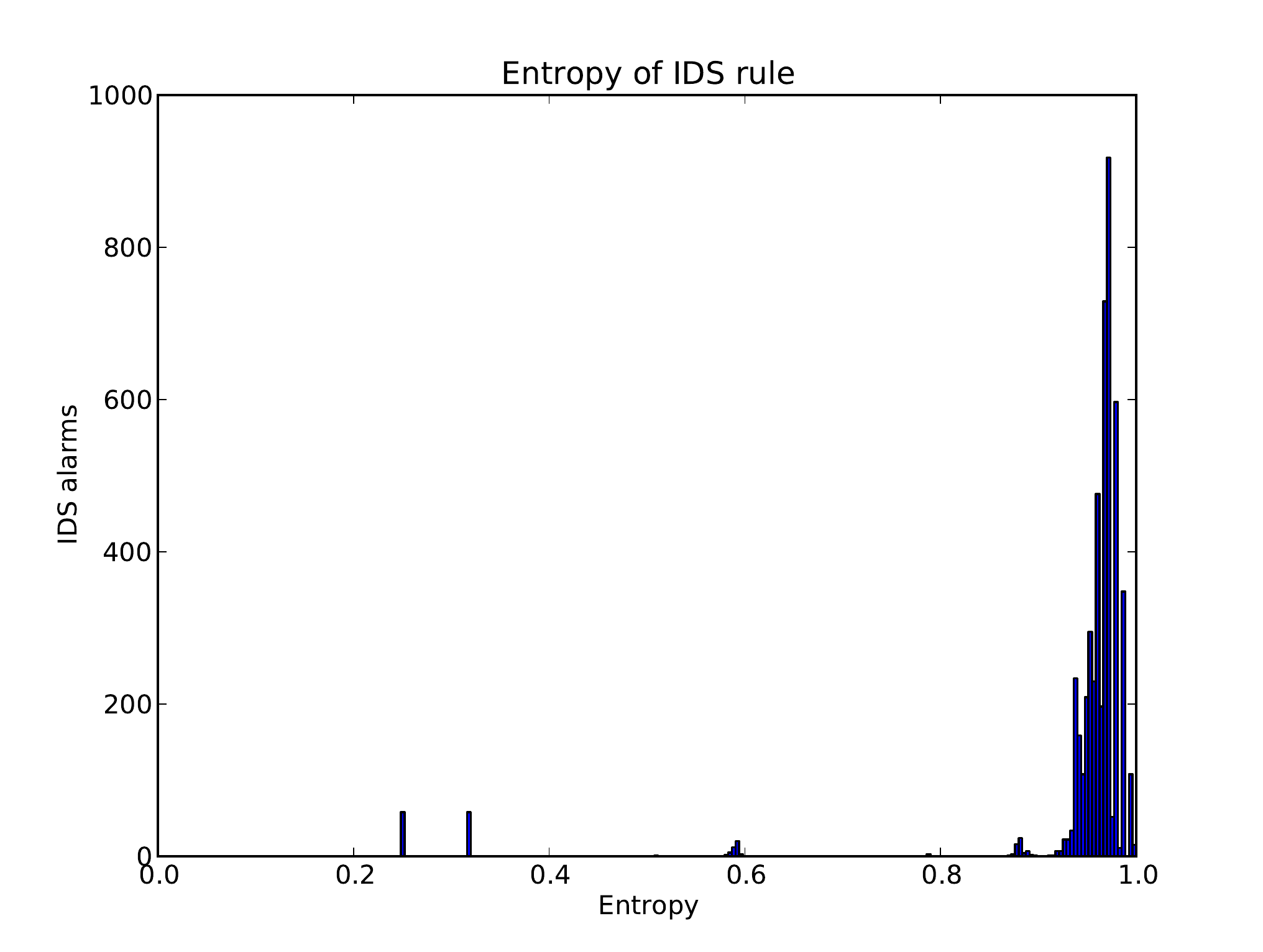}
\par\end{centering}

\caption{\label{fig:IDS-rule-1:2925}IDS rule 1:2925 1x1 GIF attempt (web bug),
illustrating a privacy leaking IDS rule.}
\end{figure}

\emph{}

Then consider a non-perfect IDS rule $R$, which in addition to the
assumed necessary and limited information leakage by the attack pattern,
also may have false alarms\index{false alarms} or other entropy sources\index{entropy sources},
as illustrated in Figure~\ref{fig:IDS-rule-1:2925}. However, it
still only detects one attack vector, that does not change between
attacks. This means that the entropy distribution function will be
unimodal\index{unimodal}, perhaps with some outliers as illustrated
in Figure~\ref{fig:IDS-rule-1:2925}. This is a simplistic model
of how an IDS rule behaves. It does not assume any particular IDS
rule implementation (e.g. whether string matching or regular expressions
are being used) and does not take any position on the type of IDS
technology being used. Experimental results have however shown that
a significant amount of all IDS rules (35-53\% in the experiments
we have performed%
\footnote{53\% of the IDS rules in the experiments performed here were unimodal,
indicating one attack vector. A former pre-experiment at a commercial
MSS provider indicated that 35\% of the IDS rules were unimodal.%
}) actually \emph{behave} in this way. However, this also means that
many IDS rules actually do \emph{not }behave this way. We will therefore
later discuss how this restriction can be removed.

\begin{figure}
\begin{centering}
\includegraphics[scale=0.55]{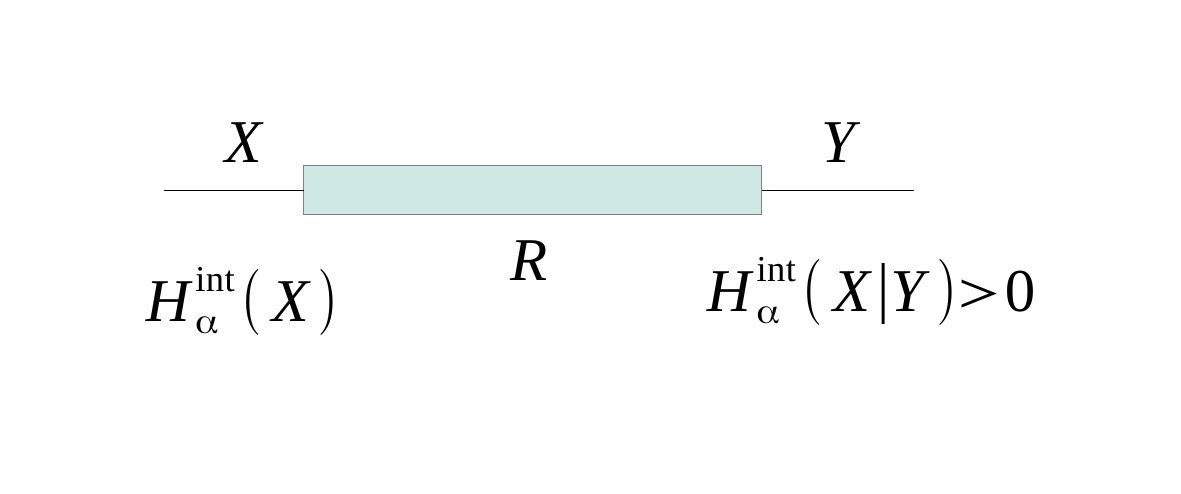}
\par\end{centering}

\caption{\label{fig:Channel-model-nonperfect-rule}Channel model of a non-perfect
IDS rule $R$ that detects a single nonmutating underlying attack
vector. $R$ may have false alarms or other entropy sources, which
means that $H_{\alpha}^{int}(X|Y)>0$.}
\end{figure}

The model of a unimodal non-perfect IDS rule is illustrated in Figure~\ref{fig:Channel-model-nonperfect-rule}.
Assume that this IDS rule generates the ordered set of $N$ IDS alarms
denoted as $Y=\{y_{1},y_{2},...,y_{N}\}$, where $P[Y=y_{i}]<P[Y=y_{j}]$
for $i<j$, $i,j\in1,2,...,N$. The inter-alarm entropy will in this
case be greater than zero for both Shannon and Min-entropy, because
${\displaystyle \sum_{i=1}^{n}P[X|Y=y_{i}]=1}$ and $P[X|Y=y_{1}]<1$.

\begin{figure}
\begin{centering}
\includegraphics[scale=0.55]{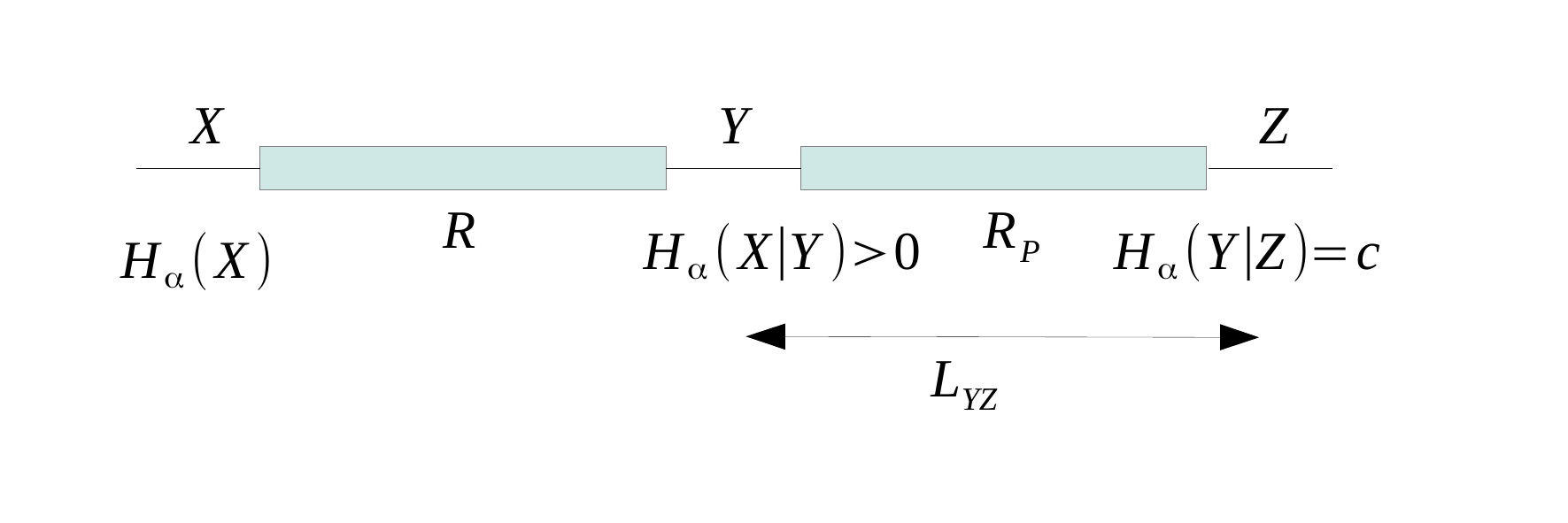}
\par\end{centering}

\caption{\label{fig:Channel-model-of-rule}Channel model of privacy leakage
from a non-perfect IDS rule $R$, measured relative to a perfect model
IDS rule.}
\end{figure}

\subsection{Privacy Leakage Model\index{privacy leakage model}}

The next question is how to model the \emph{privacy leakage} from
the non-perfect IDS rule $R$. One way to do this, is to measure the
information leakage\index{information leakage} of the non-perfect
IDS rule $R$ \emph{relative} to a perfect model IDS rule $R_{P}$,
as illustrated in Figure~\ref{fig:Channel-model-of-rule}. The communication
channel then consists of a cascade\index{cascade} of two IDS rules
(or two IDS rules connected in series), where the output of the first
IDS rule serves as input to the second IDS rule. Both IDS rules have
the objective to trigger on the same attack vector, however the first
IDS rule $R$ is non-perfect, and may have false alarms or other entropy
sources, whereas the second IDS rule $R_{P}$ is considered a perfect
model IDS rule. The advantage of using a cascading model, is that
this allows for comparing known values, and it is not dependent on
the unknown Internet traffic $X$. The set of alarms $Y$ from $R$
are known by the MSS provider and the set of expected alarms $Z$
from $R_{P}$ are also known given $Y$.

Focusing on the inter-alarm entropies is not a fruitful approach here,
since the difference in inter-alarm entropies is $H_{\alpha}^{int}(X|Y)-H_{\alpha}^{int}(Y|Z)=H_{\alpha}^{int}(X|Y)$,
because $H_{\alpha}^{int}(Y|Z)=0$. What is needed, is therefore a
measure of the limited information leakage that the perfect model
IDS rule causes. 

This initial information loss, denoted as the \emph{intra-alarm} information
loss\emph{\index{intra-alarm information loss}} $H_{\alpha}(X)$,
can be expressed by measuring the entropy of the IDS alarm in bits,
instead of measuring the inter-alarm entropy $H_{\alpha}^{int}$ (the
entropy between IDS alarms, considering the entire IDS alarm as one
symbol). The intra-alarm entropy for a perfect model IDS rule $R_{P}$
can be calculated by assuming that the IDS alarm consists of a large
sequence of bits. This can be expressed formally by considering a
given IDS alarm as $y\in\{0,1\}$ where $P[y=1]=1-P[y=0]$.

Considering the perfect model IDS rule first, then this IDS rule will
always return the same IDS alarm $Z=\{y\}$ where $y\in\{0,1\}$ with
bit-probability $\{P[y=0],P[y=1]\}$. The information leakage is defined
according to (\ref{eq:information-leakage}) as:

\begin{equation}
L_{YZ}=H_{\alpha}(X|Y=y)-H_{\alpha}(Y=y|Z=y)=H_{\alpha}(X|Y=y)\label{eq:leakage-L_YZ}
\end{equation}

Since $R_{P}$ is deterministic, then $Z$ will be determined by $Y$,
which means that $H_{\alpha}(Y|Z)=0$. Furthermore, for Shannon entropy:

\begin{equation}
H_{1}(X|Y=y)={\displaystyle \sum_{x\in\{0,1\}}}P[X=x|Y=y]log\frac{1}{P[X=x|Y=y]}
\end{equation}

$P[X=x|Y=y]=0$ for $x\neq y$, which means that this can be expressed
as:

\begin{equation}
H_{1}(X=y)={\displaystyle \sum_{y\in\{0,1\}}}P[X=y]log\frac{1}{P[X=y]}
\end{equation}

which gives:

\begin{align}
H_{1}(X & =y)=P[y=0]log\frac{1}{P[y=0]}+\\
 & (1-P[y=0])log\frac{1}{(1-P[y=0])}=c_{1}
\end{align}

This shows that $R_{P}$ has a constant privacy leakage $L_{YZ}=c_{1}$
for Shannon entropy. This can also be shown for Min-entropy by substituting
into Equation (\ref{eq:H_infty_X_given_Y}):

\begin{equation}
H_{\infty}(X|Y=y)=log\frac{1}{V(X|Y=y)}
\end{equation}

where the vulnerability $V(X|Y=y)$ can be expressed as:

\begin{equation}
V(X|Y=y)={\displaystyle \sum_{y\in\{0,1\}}P[Y=y]}{\displaystyle \max_{x\in\{0,1\}}P[X=x|Y=y]}
\end{equation}

$P[X=x|Y=y]=0$ for $x\neq y$, which means that this can be expressed
as:

\begin{equation}
V(X|Y=y)={\displaystyle P[y=0]^{2}+P[y=1]^{2}}
\end{equation}

which can be expressed as

\begin{equation}
V(X|Y=y)=1-2P[y=0](1-P[y=0])
\end{equation}

This shows that the vulnerability is $V(X|Y=y)=1$ for $P[y=0]\in\{0,1\}$.
The lowest vulnerability is $V(X|Y=y)=\frac{1}{2}$ for $P[y=0]=\frac{1}{2}$,
as expected. This means that the Min-entropy for $R_{P}$ can be expressed
as:

\begin{equation}
H_{\infty}(X|Y=y)=log\frac{1}{1-2P[y=0](1-P[y=0])}=c_{\infty}
\end{equation}

\begin{figure}
\begin{centering}
\includegraphics[scale=0.45]{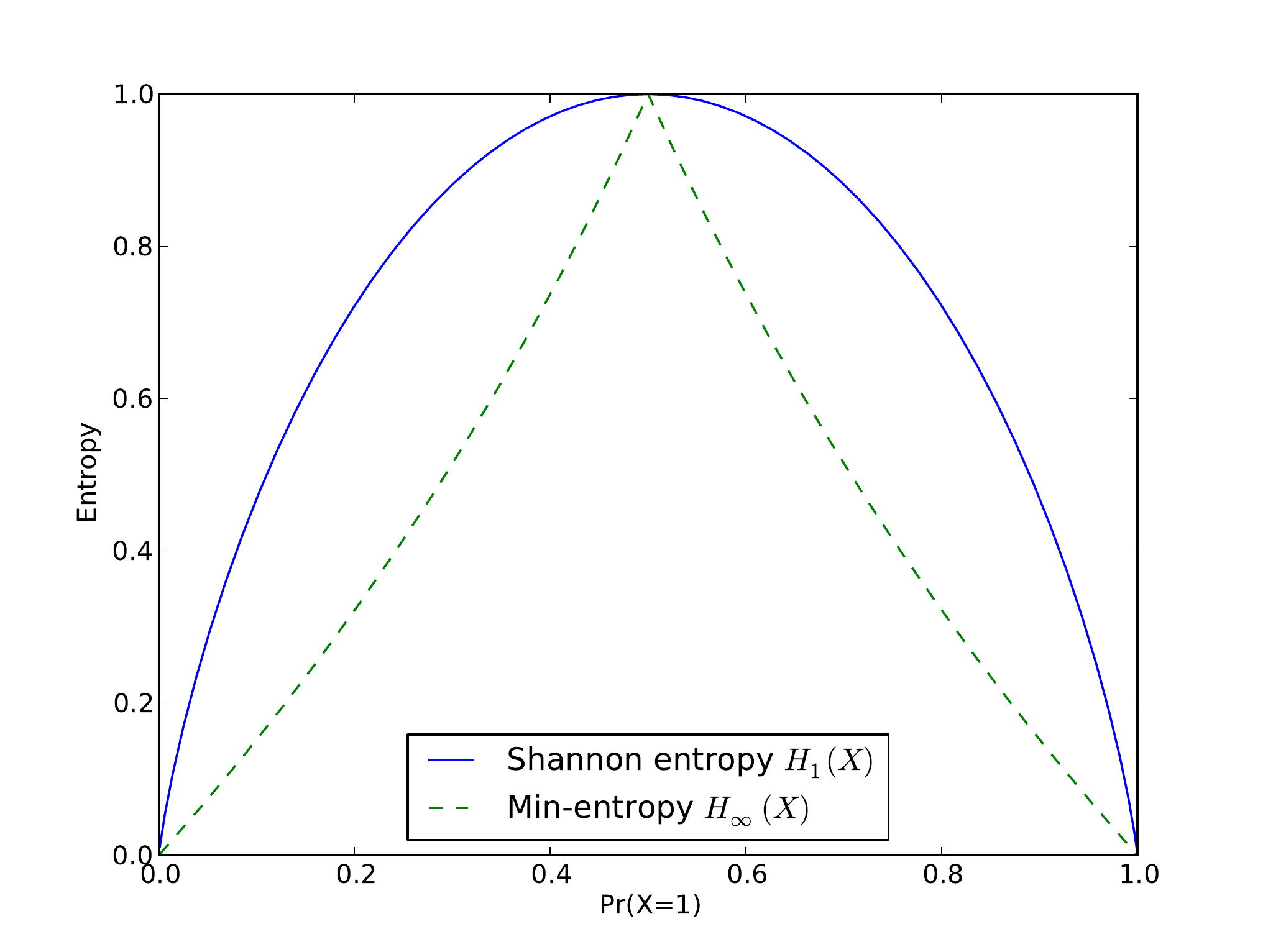}
\par\end{centering}

\caption{\label{fig:Shannon-vs.-Min-entropy}Shannon vs. Min-entropy.}
\end{figure}

This means that $R_{P}$ has a constant information leakage for both
Shannon-entropy $L_{YZ}=c_{1}$ and Min-entropy $L_{YZ}=c_{\infty}$.
However these constants are different, except in the special cases
where $P[y=0]\in\{0,\frac{1}{2},1\}$, as can be expected (see Figure
\ref{fig:Shannon-vs.-Min-entropy}). 

Let the constant information leakage for either Shannon or Min-entropy
be denoted as $c_{\alpha}$. The relative information leakage from
the IDS rule $R$ can then be formally defined as follows:
\begin{defn}
Let $R$ be a non-perfect IDS rule, that in addition to the assumed
necessary and limited information leakage by the attack pattern, also
may have false alarms or other entropy sources. Let $R_{P}$ be a
perfect model IDS rule with a limited privacy leakage $c_{\alpha}$,
$\alpha\in\{1,\infty\}$%
\footnote{It is possible to show that this definition generalises to any Rényi
entropy, however that is beyond the scope of this \paper{} , since
Min-entropy and Shannon-entropy are considered the best candidates
for the privacy leakage metric~\citep{smith_foundations_2009}.%
}. The relative information leakage\index{relative information leakage}
$L_{YZ}$ for an IDS rule $R$ with input $X$, that generates a set
of IDS alarms $Y=\{y_{1},y_{2},...,y_{N}\}$, each with probability
$P[Y=y_{i}],\, i=1,...,N$ is then defined as the difference in intra-alarm
entropy between $R$ and a perfect model IDS rule $R_{P}$ that both
trigger on the same attack vector:

\begin{equation}
L_{YZ}=H_{\alpha}(X|Y)-c_{\alpha}\label{eq:entropy-privacy-leakage}
\end{equation}

\end{defn}
If the probability distribution function\index{probability distribution function}
(PDF) of the IDS alarm entropies for a given attack vector is symmetric,
then the \emph{average} entropy\emph{\index{average entropy}} denoted
as $\overline{H_{\alpha}}(X|Y)$ for input $X$ and a sufficiently
large set of IDS alarms $Y$ can be considered as a good estimator
of $c_{\alpha}$. For skewed distributions, the \emph{median\index{median}}
may give a better estimate, given that the sample is sufficiently
large. It can furthermore be observed that the precision of this estimator
will improve with the precision of the IDS rule $R$. This means that
the information leakage of $R$ for a given IDS alarm $y_{i}$ can
be expressed as:

\begin{equation}
L_{YZ}=H_{\alpha}(X|Y=y_{i})-\overline{H_{\alpha}}(X|Y)\label{eq:L_YZ}
\end{equation}
where the average entropy can be expressed as 
\begin{equation}
\overline{H_{\alpha}}(X|Y)={\displaystyle \sum_{i=1}^{N}P[Y=y_{i}]H_{\alpha}(y_{i})}
\end{equation}
for a set of input data $X$.

\subsection{Information Leakage for a Sample of IDS Alarms}

The average entropy per byte for a \emph{sample\index{sample}} $y_{1},y_{2},...,y_{N}$
of $N$ IDS alarms generated by an IDS rule $R$ that detects a single
attack vector, can be expressed as 

\begin{equation}
\overline{H_{\alpha}}=\frac{1}{N}{\displaystyle \sum_{i=1}^{N}H_{\alpha}(y_{i})}.
\end{equation}

The information leakage for any IDS alarm $y_{j}$, denoted as $L_{R}(y_{j})$
can then be expressed as: 

\begin{equation}
L_{R}(y_{j})=H_{\alpha}(y_{j})-\frac{1}{N}{\displaystyle \sum_{i=1}^{N}H_{\alpha}(y_{i})}
\end{equation}

Further processing of the information leakage $L_{R}(y_{i})$ for
the IDS alarms $y_{1},y_{2},...,y_{n}$ can now be calculated using
traditional statistical analysis. The privacy leakage of the IDS rule
can be expressed as the standard deviation\index{standard deviation}
$\sigma_{\alpha},$ error margin\index{error margin} $2\sigma_{\alpha}$
or the 95\% confidence interval\index{confidence interval} $\pm2\sigma_{\alpha}$
of the IDS rule. This gives an indication of the expected precision
of the IDS rule. Another useful metric, is to consider the worst-case
information leakage\index{worst-case information leakage} denoted
as $L_{R}^{max}$ where $L_{R}^{max}={\displaystyle \max_{i=1}^{n}L_{R}}$,
or the minimum information leakage denoted as $L_{R}^{min}$ where
$L_{R}^{min}={\displaystyle \min_{i=1}^{n}L_{R}}$. Both of these
can be useful in statistical analyses, in addition to the standard
deviation. Furthermore, the privacy leakage can be calculated as $\pi_{R}^{L}=L_{R}\cdot I_{R}$,
where $I_{R}$ is the privacy impact estimated by the data controller.

\subsection{\label{sub:Sample-Standard-Deviation}Sample Standard Deviation of
Entropy $\sigma_{\alpha}$}

\subsubsection{Normal Distribution\index{Normal distribution}}

Assuming that the probability distribution of alarms can be approximated
using a Normal distribution, then the standard deviation\index{standard deviation}
can be calculated using the second norm\index{second norm}. 

Assume that the IDS generates a sample of \emph{$n$ }IDS alarms $(y_{1},y_{2},...,y_{N})$.
Each alarm $y_{i}$ contains payload or other potentially privacy
leaking elements or attributes from the IDS alarms generated by an
IDS rule \emph{$R$}. The sample standard deviation of the entropy
of the elements can then be expressed as:

\begin{equation}
\sigma_{\alpha}=\sqrt{\frac{1}{N-1}\sum_{i=1}^{N}(L_{R})^{2}}=\sqrt{\frac{1}{N-1}\sum_{i=1}^{N}(H_{\alpha}(y_{i})-\overline{H_{\alpha}})^{2}}
\end{equation}

\emph{ }The general properties of the \emph{variance\index{variance}}
of entropy measurements $\sigma_{\alpha}^{2}$ will fulfill the same
requirements as the standard deviation of entropy measurements. However,
the standard deviation is considered more appropriate, since it operates
with the same unit of measure as the entropy.

\subsubsection{\label{sub:Laplacian-Distribution}Laplacian Distribution\index{Laplacian distribution}}

An alternative distribution that during the experiment was shown to
fit the data well, is the Laplacian (or double exponential) distribution.
The Laplacian standard deviation, denoted as $\sigma_{\alpha}^{L}$
is based the $L^{1}$ norm (or Manhattan distance\index{Manhattan distance}),
and can be expressed as the sum of absolute deviations:

\begin{equation}
\sigma_{\alpha}^{L}=\frac{\sqrt{2}}{N}{\displaystyle \sum_{i=1}^{N}\left|H_{\alpha}(y_{i})-\overline{H_{\alpha}}\right|}
\end{equation}

A well known advantage with $\sigma_{\alpha}^{L}$, is that it will
be less influenced by outliers\index{outliers} in the tail of the
PDFs than the standard deviation of the Normal distribution.

The standard deviation of normalised entropy is a measure of the \emph{relative
information leakage\index{relative information leakage}} from an
IDS rule, under the assumption that it detects only one nonmutating
attack vector. If an IDS rule detects the attack vector perfectly
without any false alarms, then the entropy of the IDS alarms will
always be the same, and $\sigma_{\alpha}=0$. If the IDS alarm is
precise\index{precise} at detecting the attack, then only a few bits
of information will vary between IDS alarms. This means that all alarms
will have similar entropy with low standard deviation and therefore
also low information leakage. However if the IDS rule also has a significant
amount of false alarms, or gets entropy from other sources then the
entropy variance, and therefore also the information leakage from
the IDS rule, will increase.

\subsection{Aggregating $\sigma_{\alpha}$}

This subsection shows how the standard deviation of entropy metric
can be aggregated for a set of IDS rules. Assume that an IDS uses
a rule set denoted as $R_{all}$ with $m$ IDS rules $R_{all}=\{R_{1},R_{2},...,R_{m}\}$.
Each IDS rule $R_{i}$ matches independently a set of $N_{i}$ IDS
alarms:\\
 $Y_{i}=\{y_{i,1},y_{i,2},...,y_{i,N_{i}}\}$, $i=1,2,...,m$ where
the number of IDS alarms $N_{i}$ typically will vary between IDS
rules. Furthermore, assume that the IDS alarms are independent and
non-overlapping, i.e. $Y_{i}\cap Y_{j}=\emptyset$ for $i\neq j$.
This means that all IDS alarms, denoted $Y_{all}$, can be expressed
as $Y_{all}={\displaystyle \bigcup_{i=1}^{m}}Y_{i}$.

Assume that an IDS rule $R_{i}$ has entropy standard deviation denoted
as $\sigma_{i}$ and resulting standard deviation denoted as $\sigma_{all}$.
The aggregated metric\index{aggregated metric} should furthermore
fulfill the following criteria in order to provide meaningful aggregation:
\begin{lyxlist}{00.00.0000}
\item [{C1}] If all IDS rules have the same standard deviation, say $\sigma_{i}$,
then $\sigma_{all}$ should also be the same, i.e. $\sigma_{all}=\sigma_{i}$.
\item [{C2}] The resulting entropy standard deviation should be weighted
according to how many alarms that trigger on a given IDS rule $R_{i}$.
\end{lyxlist}
 Each IDS rule should be assessed individually, in the same way as
each underlying vulnerability should be assessed individually. This
means that a \emph{weighted average\index{weighted average}}, weighted
by number of alarms from each IDS rule, can be used as aggregation
function\index{aggregation function} for $\sigma_{all}$, i.e:

\begin{equation}
\sigma_{all}=\frac{{\displaystyle {\displaystyle \sum_{i=1}^{m}N_{i}\sigma_{i}}}}{{\displaystyle \sum_{i=1}^{m}N_{i}}}\label{eq:sigma_all}
\end{equation}

This function fulfills criterion C1, since the resulting average weighted
sum is the same if $\sigma_{i}$ is the same for all IDS rules $R_{i}$
and it fulfills C2 by weighting the standard deviation against number
of IDS alarms.

\subsection{\label{sub:Several-Attack-Vectors}IDS Rules Detecting Several Attack
Vectors\index{detecting several attack vectors}}

\begin{figure}
\begin{centering}
\includegraphics[scale=0.35]{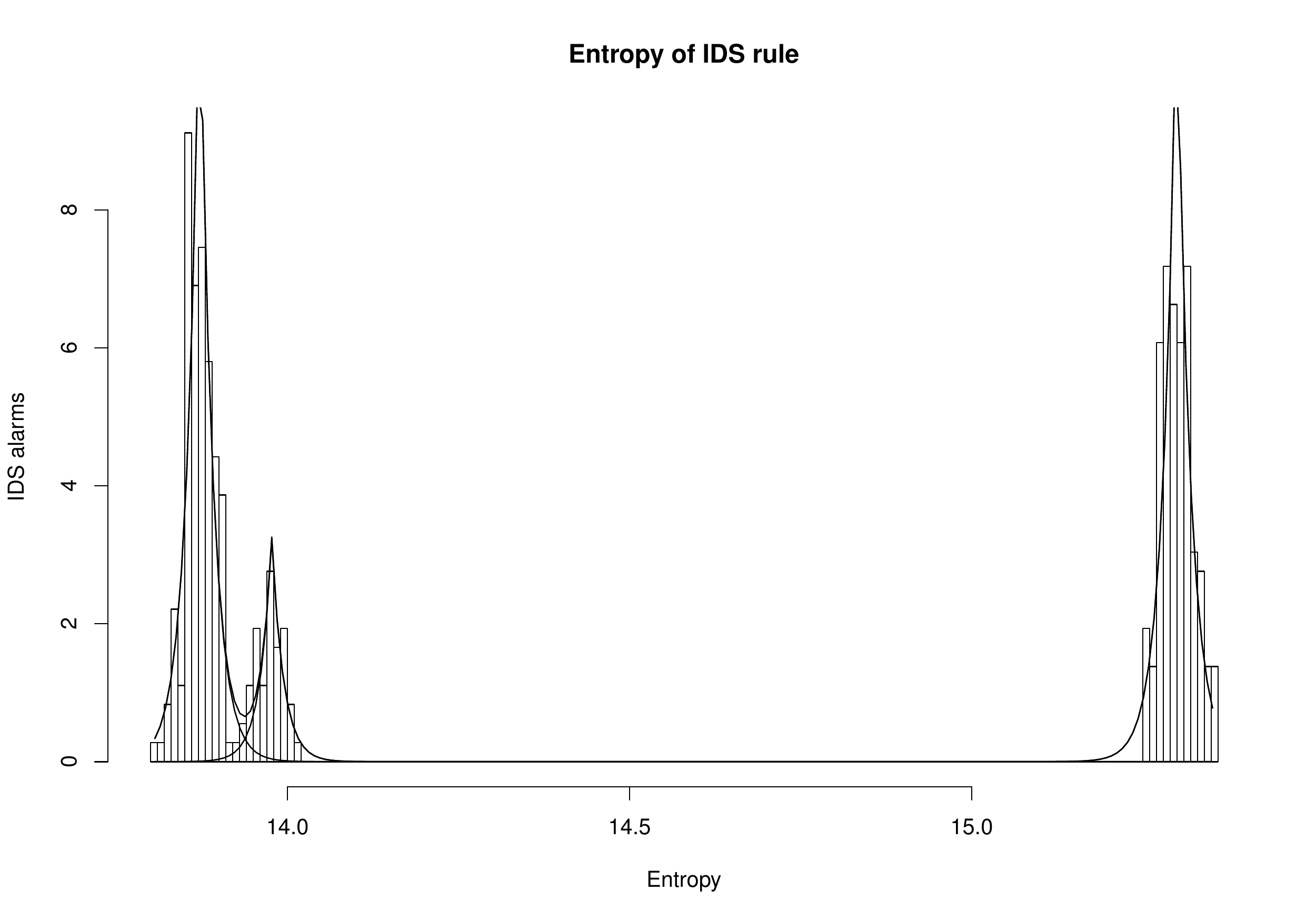}
\par\end{centering}

\caption{\label{fig:Multimodal-rule}Payload length corrected Shannon octet-entropy
distribution of IDS rule (Snort SID 1:11969) matching three attack
vectors.}
\end{figure}

A significant part of the IDS rules will detect more than one attack
vector, as illustrated in Figure~\ref{fig:Number-of-attack-vectors}.
The data set used in this \paper{}  has 47\% of the IDS rules with
more than one attack vector. An earlier preliminary experiment at
a commercial MSS provider shows even higher percentage (65\%). An
indication of an IDS rule that detects several attack vectors, is
that the entropy probability distribution is multi-modal\index{multi-modal}.
Figure \ref{fig:Multimodal-rule} shows an example IDS rule that matches
three privacy leaking attack vectors. The Figure shows the payload
entropy distribution of the Snort IDS rule with SID 1:11969 VOIP-SIP
inbound 401 Unauthorized. A payload length correction causes the metric
to be larger than one, and is required to make the metric incentive
compatible%
\footnote{Incentive compatibility \textendash{} a characteristic of mechanisms
whereby each agent knows that his best strategy is to follow the rules,
no matter what the other agents will do~\citep{durlauf_incentive_2008}.%
}. The details of this can be ignored for now, since this will be discussed
in Section \ref{sub:Payload-Length-Correction}. Each attack vector
cluster corresponds to a different SIP service provider.

A clustering algorithm\index{clustering algorithm} is needed to identify
each underlying attack vector for multi-modal distributions. Each
individual cluster will in this case represent an attack vector, which
behaves in a similar way as a non-perfect IDS rule described in Section~\ref{sub:A-Non-perfect-IDS-rule}.
This means that the privacy leakage of each attack vector cluster
can be calculated as the entropy standard deviation over all samples
belonging to the cluster, and the resulting privacy leakage for the
IDS rule can be calculated by aggregating the data over all IDS rules
in the cluster using Equation \ref{eq:sigma_all}.

\subsection{How to Perform the Clustering}

There are two main types of clustering algorithms: hard clustering\index{hard clustering}
and soft clustering\index{soft clustering}. Hard clustering algorithms
assign each sample to a given cluster. Examples of a hard clustering
algorithm is the popular k-means and k-medians algorithms~\citep{macqueen_methods_1967,bradley_clustering_1997}.
Hard clustering is however not appropriate for clustering the IDS
rules, since it cuts off the samples at the tail of the distribution
where two distributions overlap. This will give a bias towards lower
entropy standard deviation than can be expected.

Soft clustering is then a better approach, since it assigns the probability
that each sample belongs to a given cluster, instead of assigning
each sample to a given cluster. A commonly used soft clustering technique
is the Expectation Maximisation (EM) algorithm\index{Expectation Maximisation}~\citep{dempster_maximum_1977}.
This soft-clustering method provides a Maximum Likelihood estimate
of the underlying data distribution as a mixture of assumed probability
distributions. The EM-algorithm is basically a two-step hill-climbing
technique where the first step (E-step) calculates the expectation
\index{expectation}of the log-likelihood\index{log-likelihood} using
the current estimate of the parameters of the underlying probability
distributions. The second step (M-step) computes the parameters that
maximise the expected log-likelihood identified during the E-step. 

There are however some drawbacks with the EM-algorithm\index{EM-algorithm}.
It is prone to get stuck in local minima, which means that it is sensitive
to the initial cluster parameters. We use the cluster centers identified
by k-means, since this is a generally recommended method of initialising
the cluster centers%
\footnote{We used k-means from the Python module scikit-learn to initialise
the EM algorithm \citep{scikit-learn}.%
}. Another issue is the selection of number of clusters. Too many clusters
may cause EM to overfit\index{overfit} the data, whereas too few
clusters may give a poor representation of the distribution of the
samples.  

It is commonly assumed that the underlying probability distribution
either is a mixture of Gaussian\index{Gaussian} or Laplacian\index{Laplacian}
probability density functions. Both outliers\index{outliers} and
skewedness\index{skewedness} have been found to be significant during
the experimental analysis in Section \ref{sub:Analysis-of-Alarm-PDF}.
We have therefore decided to model the probability distribution as
a mixture of Laplacian probability density functions using the method
proposed in \citep{cord_feature_2006}. This method is based on order
statistics\index{order statistics} (uses a weighted median instead
of the mean), and is therefore more robust against outliers and skewedness
than using a Gaussian mixture~\citep{cord_feature_2006}. The remainder
of this section highlights the necessary theory and notation to understand
how we have implemented the Laplacian mixture based clustering.

\subsection{Laplacian Mixture Model\index{Laplacian Mixture Model}}

This section defines the general notation, which is based on the well-known
theory of learning finite mixture models~\citep{bailey_fitting_1994,figueiredo_unsupervised_2002}.
Furthermore, the Laplacian Mixture Model used here, is based on \citep{cord_feature_2006}.
Our implementation is simplified compared to the original solution,
since only univariate clustering is needed. Let $\mathscr{H}_{R}$
be a random variable representing the IDS alarm entropies of an IDS
rule $R$, with $H_{\alpha}$ representing one particular outcome
of $\mathscr{H}_{R}$. This random variable is expressed as:

\begin{equation}
P[\mathscr{H}_{R}=H_{\alpha}|\Theta]={\displaystyle \sum_{k=1}^{K}\beta_{k}P[\mathscr{H}_{R}=H_{\alpha}|\Theta=\theta_{k}]}
\end{equation}

where $\beta_{1},...,\beta_{K}$ are the mixing probabilities, each
$\theta_{k}$ is the set of parameters defining the $k$-th component
of the mixture and $\Theta=\{\theta_{1},...,\theta_{K},\beta_{1},...,\beta_{K}\}$
is the complete set of parameters that define the mixture. Being probabilities,
$\beta_{k}$ must satisfy $\beta_{k}\geq0$ and ${\displaystyle \sum_{k=1}^{K}}\beta_{k}=1$.
It is assumed that all the components of the mixture are Laplacian
distributions $P[\mathscr{H}_{R}=H_{\alpha}|\theta_{k}]=\mathscr{L}(H_{\alpha}|\theta_{k}=(\tilde{\mu},\lambda))$.
The Laplacian distribution is defined as:

\begin{equation}
\mathscr{L}(H_{\alpha}|\tilde{\mu_{k}},\lambda_{k})=\frac{1}{2\lambda_{k}}exp\left(-\frac{\left|H_{\alpha}-\tilde{\mu}_{k}\right|}{\lambda_{k}}\right)
\end{equation}

where $H_{\alpha}(y_{i})$ is the entropy of the IDS alarm $y_{i}$,
$\lambda_{k}>0$ is the scale parameter and $\tilde{\mu}_{k}$ is
the median for mixture component $\theta_{k}$. In the remainder, assume
the shorthand notation that $\mathscr{L}_{\alpha,i,k}=\mathscr{L}(H_{\alpha}(y_{i})|\theta_{k})$.

\subsection{EM-Algorithm for Laplacian Mixture Model}

The implementation of the EM-algorithm is based on \citep{cord_feature_2006,figueiredo_unsupervised_2002}.
Assume that the EM-algorithm is performing cluster analysis on a \emph{sample\index{sample}}
of $N$ ordered entropy values $\mathbf{H_{\alpha}}=(H_{\alpha,}(y_{1}),H_{\alpha}(y_{2}),...,H_{\alpha}(y_{N}))$,
where $H_{\alpha}(y_{i})<H_{\alpha}(y_{j})$ for $i<j$, $i,j\in1,2,...,N$.
These entropy values are calculated over the IDS alarms $y_{1},y_{2},...,y_{N}$
generated by an IDS rule $R$. The Expectation Maximisation algorithm
for the Laplacian Mixture Model then consists of two steps that are
iterated until convergence is detected:

\emph{E-step:} calculate the conditional expectation of the complete
log-likelihood $w_{i,k}=log(P[\mathscr{H}_{R}=H_{\alpha}(y_{i})|\Theta=\theta_{k}])$
that $H_{\alpha}(y_{i})$ comes from the $k$-th component of the
mixture:

\begin{equation}
w_{i,k}=\frac{\beta_{k}\mathscr{L}_{\alpha,i,k}}{\sum_{k=1}^{K}\beta_{k}\mathscr{L}_{\alpha,i,k}}
\end{equation}

\emph{M-step:} estimate new model parameters $\theta_{k}=(\tilde{\mu}_{k},\lambda_{k})$
and weights $\beta_{k}$ that maximise the log-likelihood $log(\mathscr{L}_{\alpha,i,k})$
of the model:

\begin{eqnarray}
\tilde{\mu}_{k} & = & wmedian(\mathbf{H_{\alpha}},k)\\
\lambda_{k} & = & \frac{1}{\sum_{i=1}^{N}w_{i,k}}\sum_{i=1}^{N}w_{i,k}\left|H_{\alpha}(y_{i})-\tilde{\mu}_{k}\right|\\
\beta_{k} & = & \frac{{\displaystyle \sum_{i=1}^{N}}w_{i,k}}{{\displaystyle \sum_{i=1}^{N}}{\displaystyle \sum_{k=1}^{K}}w_{i,k}}
\end{eqnarray}

where the algorithm to calculate the weighted median\index{weighted median}
for a given cluster $k$, according to \citep{cord_feature_2006},
is described in Algorithm~\ref{alg:Weighted-median}.

\begin{algorithm}
\begin{algorithmic}[1]
	\Function{wmedian}{$\mathbf{H_{\alpha}},k$}
		\State $Q=(q_0=0,q_1=0,...,q_N=0)$
		\State sum=0
		\For {$i \leftarrow 1,...,N $}
			\State $sum \leftarrow sum + w_{i,k}$
			\State $q_i=sum$
		\EndFor
		\For {$i \leftarrow 1,...,N $}
			\If {$q_i>\frac{1}{2}q_N>q_{i-1}$}
				\State return ($H_\alpha(y_i)+H_\alpha(y_{i-1}))/2$
			\ElsIf {$q_i=\frac{1}{2}q_N$}
				\State return $H_\alpha(y_i)$
			\EndIf
		\EndFor
	\EndFunction
\end{algorithmic}

\caption{\label{alg:Weighted-median}Weighted median.}
\end{algorithm}

The algorithm uses the Minimum Message Length\index{Minimum Message Length}
(MML) as stop criterion\index{stop criterion} \citep{wallace_information_1968},
assuming one-dimensional data. We do not go into details on the MML
criterion and just present the implemented solution here. The detailed
derivation of the MML criterion used can be found in \citep{figueiredo_unsupervised_2002}.

\begin{eqnarray}
MML & = & {\displaystyle \sum_{k=0}^{K}log(\frac{N\beta_{k}}{12})+\frac{K}{2}log\frac{N}{12}+\frac{3K}{2}-}\nonumber \\
 &  & max_{k}\left\{ {\displaystyle \sum_{i=1}^{N}}log{\displaystyle (w_{i,k})}\right\} \label{eq:MML}
\end{eqnarray}

The last term of Equation \ref{eq:MML} is derived from the fact that
the minimum of the $MML$ criterion over $\Theta$ can be obtained
by using the negative maximum of the log-likelihood (the last term),
since

\begin{equation}
max_{\Theta}\left\{ log\left({\displaystyle P[\mathscr{H}_{R}|\Theta]}\right)\right\} =max_{k}\left\{ {\displaystyle \sum_{i=1}^{N}}log{\displaystyle (w_{i,k})}\right\} .
\end{equation}
 The algorithm stops when the difference in MML length between two
iterations is less than $\epsilon_{MML}=1\times10^{-4}$. In addition
to the MML criterion, the implementation of the EM algorithm requires
at least 40 iterations to converge initially, and at least 20 iterations
to converge after modifications of the cluster definitions. This is
to avoid accidentally hitting a local MML minimum before convergence
has occurred.

\subsection{Determining the Optimal Number of Clusters\index{optimal number of clusters}
$k$}

We initially tested the method for estimating the number of components
in \citep{figueiredo_unsupervised_2002}. This method worked for nice
continuous distributions, however it did not work equally well for
for noisy or a mixture containing binomial distributions, since the
EM-algorithm then easily got stuck in local modes. Overfitting was
also a significant problem for binomial distributions.

Furthermore, to judge whether a cluster should be interpreted as an
attack vector or not typically requires that the data controller does
some investigation of the IDS alarms. This means that some degree
of manual intervention typically will be required during the clustering
to assert obvious clusters that the clustering algorithm has missed
or delete clusters where overfitting occurs. A typical example of
overfitting is where several components with the same median are used
to represent a given cluster. Another example is for skewed distributions,
where the EM attempts to fit the skewed curve by overfitting the data.

We implemented a simple user interface for managing the clusters.
It supports configuration of the initial number of clusters $k$ as
well as managing the model definition $\Theta$ after the initial
configuration. The program also supports selecting type of entropy
data and IDS rule to analyse from the datasets. The user interface
for managing the clustering consists of the following functions:
\begin{lyxlist}{00.00.0000}
\item [{\emph{setcl($k$,$\tilde{\mu}_{k}$)\index{setcl()}}}] Assert
that the cluster number \emph{$k$} has a mode at $\tilde{\mu}_{k}$.
\item [{\emph{delcl(clusterlist)\index{delcl()}}}] Delete clusters at
index \emph{clusterlist}. Deleted clusters are marked with $\theta_{k}=(\mu_{k}=0,\rho_{k}=0,\beta=0)$.
\item [{\emph{pickcl()\index{pickcl()}}}] Pick the cluster to be asserted
by clicking the mouse at the position to be asserted in the histogram
showing the frequency distribution of the IDS alarm entropies. If
there are no clusters that are marked as deleted, then the least significant
cluster (with lowest $\beta_{k}$) will be chosen.
\end{lyxlist}
 After having modified the clusters, the EM-algorithm continues by
typing the \emph{cont} command in the debugger. When the data controller
is satisfied with the cluster definition, typing \emph{cont} without
modifying the cluster causes the algorithm to finish and print out
the calculated privacy leakage for each cluster and also the aggregated
privacy leakage for the IDS rule $R$.

\subsection{Calculating the Privacy Leakage for Clusters}

The privacy leakage\index{privacy leakage} for the identified clusters
is calculated after the data controller has asserted that the relevant
clusters have been identified and that the EM-algorithm subsequently
has converged. All probability mass is then assigned to the clusters,
which means that the privacy leakage can be calculated for the given
IDS rule $R$.

First, the model $\Theta$ will in itself give an indication of the
privacy leakage in the form of the entropy standard deviation of the
Laplacian function $\mathscr{L}(H_{\alpha}(y_{i})|\theta_{k})$ for
a given cluster $k$. It is a well known fact that this can be calculated
from the scale parameter $\lambda_{k}$ for a Laplacian distribution
as $\sigma_{k}^{\mathscr{L}}=\sqrt{2}\lambda_{k}$. However to be
able to aggregate the entropy standard deviation over all clusters,
the relative proportion of the samples for a given cluster $\theta_{k}$
must be estimated, which is exactly what $\beta_{k}$ indicates. This
means that the resulting entropy standard deviation for the IDS rule
$R$ can be calculated as the weighted average using Equation \ref{eq:sigma_all},
substituting $N_{i}$ with $\beta_{k}$: 

\begin{equation}
\sigma_{R}^{\mathscr{L}}={\displaystyle {\displaystyle \sum_{k=1}^{K}\beta_{k}\sigma_{k}^{\mathscr{L}}}}.\label{eq:stdev_pr_rule}
\end{equation}

A disadvantage by using $\sigma_{k}^{\mathscr{L}}$, is that this
only will be correct if the model fits the data reasonably well. This
may be true in some cases, however the sample distributions in the
experiments do in several cases deviate significantly from the model
due to outliers, heavy tails or noise. In these cases, it will be
more correct to have a measure of $\sigma_{\alpha}$ that is based
on the underlying samples $H_{\alpha}(y_{i})$ weighted according
to the conditional expectation $w_{i,k}$ of the model distributions
defined by $\Theta$, so that the weighted entropy is described by
$w_{i,k}H_{\alpha}(y_{i})$. This means that the model distributions
is used to specify how the samples are divided between the clusters,
instead of defining the clusters directly. The mean value of the cluster
entropies for cluster \emph{k} can then be expressed as: 
\begin{equation}
\mu_{k}=\frac{{\displaystyle \sum_{i=1}^{N}w_{i,k}H_{\alpha}(y_{i})}}{{\displaystyle \sum_{i=1}^{N}w_{i,k}}}
\end{equation}

and the Normal standard deviation can be expressed in a similar way
as:

\begin{equation}
\sigma_{k}=\sqrt{\frac{{\displaystyle \sum_{i=1}^{N}w_{i,k}\left(H_{\alpha}(y_{i})-\mu_{k}\right)^{2}}}{{\displaystyle \sum_{i=1}^{N}w_{i,k}}}}.
\end{equation}

Furthermore, the Laplacian standard deviation, based on the $L^{1}$
norm, can be expressed in terms of the conditional expectation $w_{i,k}$
and the median of the mixture component $\tilde{\mu}_{k}$ as:

\begin{equation}
\sigma_{k}^{L}=\sqrt{2}\frac{{\displaystyle \sum_{i=1}^{N}w_{i,k}\left|H_{\alpha}(y_{i})-\tilde{\mu}_{k}\right|}}{{\displaystyle \sum_{i=1}^{N}w_{i,k}}}.
\end{equation}

The resulting aggregated entropy standard deviation for the IDS rule
$R$ can in both these cases be calculated from Equation \ref{eq:stdev_pr_rule}
by substituting the relevant standard deviation into the equation.
The clustering analysis tool prints out both the individual standard
deviations per cluster as well as the resulting standard deviation
for the IDS rule based on both the standard deviation of the model
$\sigma_{R}^{\mathscr{L}}$, Normal standard deviation $\sigma_{k}$
and Laplacian standard deviation $\sigma_{k}^{L}$. It is useful to
compare these, since a large deviation between $\sigma_{R}^{\mathscr{L}}$
and the other standard deviations indicate a poor model fit, which
may or may not be relevant depending on examination of the underlying
data. 

One can for example expect good model fit\index{model fit} for IDS
rules with some Gaussian or Laplacian noise, since this is close to
the expected model of privacy leakage. However very noisy rules that
match random traffic will get a poor model fit. An example of this
is the IDS rule 1:1394000 in our experiments that detects random traffic.
It has a standard deviation over all data of 6.7 for both Normal and
Laplacian standard deviation, but only a model standard deviation
of $\sigma_{1:1394000}^{\mathscr{L}}$=1,44 . In such cases the standard
deviation of the model $\sigma_{R}^{\mathscr{L}}$ will not be usable.
Another example is if $\sigma_{k}$ is significantly larger than $\sigma_{k}^{L}$,
then $\sigma_{k}$ may be unduly influenced by outliers, which means
that $\sigma_{k}^{L}$ would be the more robust estimate. In general,
the Laplacian standard deviation can be expected to give the most
conservative estimate, which is least influenced by skewedness and
outliers.

\subsection{Summary of EM-based Clustering}

The Laplacian Mixture Model is implemented using the EM-algorithm.
A semiautomatic process is used to identify the underlying clusters
in the IDS alarms. The standard deviation of entropy metric is then
calculated for each cluster and also the aggregated metric for the
entire IDS rule. A possible attack on the clustering method, is an
overfitting attack\index{overfitting attack} where a MSS provider
decides to shirk by deliberately overfitting the attack vectors, by
asserting too many clusters during the clustering process. It is therefore
important that the role as data controller is separate from the role
as security manager, and also that external quality assurance entities
like certification organisations oversee the operation, to ensure
that it is not overly privacy invasive. It must be emphasised that
the objective not necessarily is to match the underlying probability
distribution as closely as possible. The objective is rather to identify
any likely attack vectors, and distribute the samples between these.
The EM algorithm does this reasonably well.

The EM-based clustering generalises the privacy leakage metric to
work for IDS rules that detect more than one attack vector. This generalisation
is necessary, since our experiments have shown that a significant
amount of all IDS rules trigger on more than one underlying attack
vector. An advantage with this generalisation, is that it avoids the
incentive incompatibility of the single cluster metric, which would
encourage a shirking MSS provider to cheat by splitting up IDS rules
into smaller IDS rules detecting a single attack vector.

\section{\label{sec:Detailed-Analysis-of}Detailed Analysis of $\sigma_{\alpha}$}

This section does a more thorough investigation of the standard deviation
of entropy\index{standard deviation of entropy} metric $\sigma_{\alpha}$.
The objective of this discussion is to do an analysis of the convergence
speed\index{convergence speed} required to reliably detect random
uniform input data as a function of the data length. It is expected
that random uniform input data converges towards zero entropy standard
deviation for a sufficiently long data series. This convergence speed
is an important decision factor for the selection of entropy algorithm
and symbol definition\index{symbol definition}, since the IDS alarm
entropies are calculated over a limited number of IDS alarms. Furthermore,
it is discussed which metric and symbol definition that works best
for distinguishing between plaintext and encrypted data. This analysis
shows which entropy type (Min- or Shannon entropy) and symbol size\index{symbol size}
(bit or octet) that is best for calculating privacy leakage in IDS
rules.

\subsection{\label{sub:How-Should-Entropy}Entropy Calculation}

There are at least three obvious ways of selecting the symbol space
that is used to calculate the entropies:
\begin{enumerate}
\item Define the payload of the IDS alarm as the symbol, i.e. calculate
the \emph{inter-alarm} entropy\emph{\index{inter-alarm entropy}};
\item Use binary entropy, i.e. the \emph{intra-alarm} entropy\emph{\index{intra-alarm entropy}}
as described in Section \ref{sub:How-Should-Entropy};
\item Use octets\index{octets}, i.e. 8-bit words, which commonly are used
to define the character set in computer systems.
\end{enumerate}
Other word sizes are possible, however these are considered the most
common and interesting ones for our purpose. Each of these symbol
definitions have their advantages and disadvantages, and it is important
to note that the entropy values calculated from each of these definitions
typically will be different. It has already been shown that the intra-alarm
entropy calculated from bit-entropy is different from the inter-alarm
entropy by a constant value. Furthermore, the inter-alarm entropy
is not possible to use, since it can not be used to calculate the
standard deviation of entropy.

Bit-entropy\index{bit-entropy} was used to develop the Equation \ref{eq:L_YZ},
since it is the easiest way to develop the theory for the privacy
leakage metric. The entropy standard deviation formula is however
not dependent on any particular symbol definition, as long as the
symbol definition ensures that the entropy standard deviation in the
worst case, i.e. for random, uniform data, can be measured to be sufficiently
close to zero for encrypted traffic. It is assumed that $\sigma_{\alpha}$
converges towards zero for random, uniform data as a function of input
data length, however the convergence speed is unknown and must be
investigated. It can furthermore be observed that for a perfect encryption
scheme that is approximated by random uniform data, the symbol definition
does not matter, since random uniform data does not leak any information.
This means that if the objective is to purely detect whether the information
conveyed is encrypted or not, then the entropy scheme with fastest
convergence speed may make sense to use.

This means that the minimum length of data required to reliably detect
that random uniform data has zero variance (i.e. speed of convergence)
is an important design factor that this metric relies on. It can be
expected that different entropy metrics will have different convergence
speed\index{convergence speed}. In particular, can Min-entropy be
expected to converge more slowly, since it only considers the maximum
symbol occurrence probability, and not a weighted sum of all symbol
occurrence probabilities, as Shannon entropy does.

\subsection{Entropy Bias\index{entropy bias} of Finite Length Encrypted Data}

A question that needs to be investigated, is therefore how different
entropy standard deviation metrics $\sigma_{\alpha}$ (Shannon- or
Min-entropy) respond to random uniform data strings of varying length,
and also how it is influenced by the symbol width, i.e. whether bit-entropy
or octet-based entropy is used. The reason for this, as discussed
in Subsection \ref{sub:Laplacian-Distribution}, is that the metric
shall be able to measure privacy leakage sufficiently close to zero
in the following three cases:
\begin{enumerate}
\item For a perfect model IDS rule\index{perfect model IDS rule} $R_{P}$
which detects and displays one or more non-changing attack vectors
perfectly;
\item for anonymised\index{anonymised} IDS alarms from the IDS rule;
\item as a limit case for encrypted\index{encrypted} (e.g. pseudonymised)
IDS alarms from the IDS rule, as the number of bits $n$ in the IDS
alarm goes towards infinity.
\end{enumerate}

The entropy standard deviation bias for finite length encrypted data,
denoted as $\sigma_{\alpha}^{bias}$, can be analysed by simulating
the response function of $\sigma_{\alpha}^{bias}$ as a function of
number of bits of data. The simulation is based on a set of Monte-Carlo
experiments\index{Monte-Carlo experiment}, one for each octet of
data. Each standard deviation is the average of an ensemble of 10000
experiments. Bit-length is calculated for each octet as eight times
the octet length, in order to have comparable x-axis values for bit-
and octet-based data. The experiments are based on simulations using
random uniform data selection, which means that a Normal distribution
can be assumed.

\begin{figure}
\begin{centering}
\includegraphics[scale=0.45]{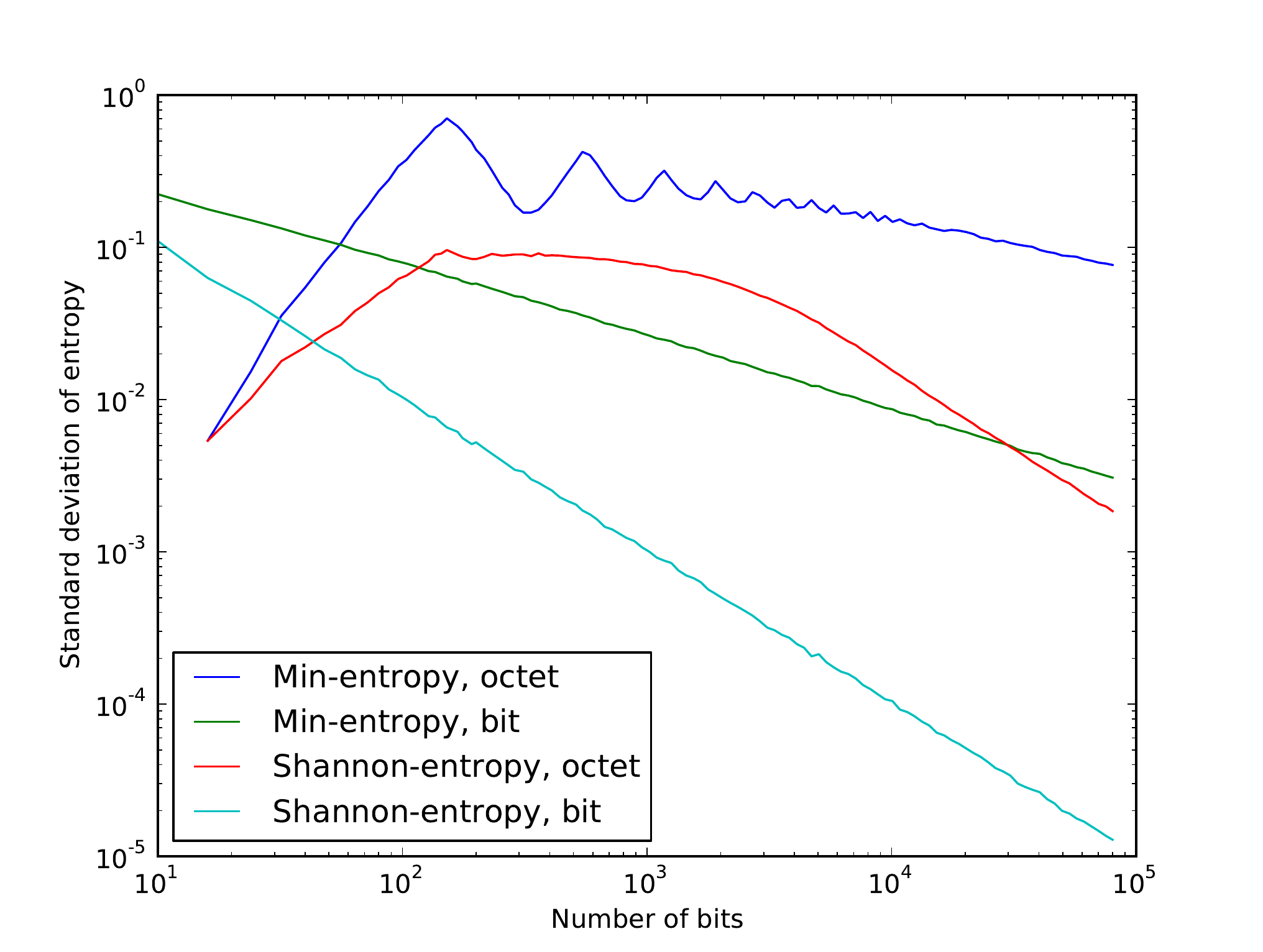}
\par\end{centering}

\caption{\label{fig:Log-log-plot-of-entropy-std}Log-log plot of entropy standard
deviation as a function of number of bits input data for Min and Shannon
entropy and bit and octet symbol definition.}
\end{figure}

Figure \ref{fig:Log-log-plot-of-entropy-std} shows a log-log plot\index{log-log plot}
of the entropy standard deviations. The bit-entropies both appear
to be log-linear\index{log-linear}, which means that the bias for
detecting a perfectly encrypted IDS alarm with length $n$ bits can
be expressed as $log_{2}(\sigma_{\alpha}^{bias})=log_{2}(\gamma_{\alpha}+\psi_{\alpha}n)$,
where $\gamma_{\alpha}$ is the offset and $\psi_{\alpha}$ is the
slope of the log-log scale. This gives $\sigma_{\alpha}^{bias}=2^{\gamma_{\alpha}}n^{\psi_{\alpha}}$,
where $2^{\gamma_{\alpha}}$ is constant. The slope can be calculated
from the experimental data, which shows that that $\psi_{1}=-1.005\approx-1$
for Shannon bit-entropy and $\psi_{\infty}=-0.479\approx-\frac{1}{2}$
for Min-entropy. This means that $\sigma_{1}^{bias}\approx\frac{2^{\gamma_{_{1}}}}{n}$,
whereas $\sigma_{\infty}^{bias}\approx\frac{2^{\gamma_{\infty}}}{\sqrt{n}}$,
which means that Shannon bit-entropy converges by an order of $O(n^{-\frac{1}{2}})$
faster towards zero than Min-entropy%
\footnote{This means that each factor in the bit-entropy calculations (one for
Min-entropy and two for Shannon entropy) contributes with a convergence
speed of $O(n^{-\frac{1}{2}}).$%
}. Shannon bit-entropy has initially 2.7 times less bias\index{bias}
than Min-entropy for perfectly encrypted (i.e. random uniform) data.
\emph{}

The octet-based entropies\index{octet-based entropies} perform very
poorly during the initial transient phase, but are then stabilised
on a slope similar to the respective bit-entropy slopes, as shown
in Figure \ref{fig:Log-log-plot-of-entropy-std}. This means that
there is a significant, but approximately constant, difference between
the bit- and octet-based metrics after the initial transient phase.
Shannon bit-entropy\index{bit-entropy} entropy ends up with a precision
143 times better than Shannon octet-entropy after 80 kbit. The difference
in precision between bit- and octet-based Min-entropy is smaller,
only 25 times.

A nice property is that the bias\index{bias} is systematic, which
means that the entropy standard deviation calculations may be able
to compensate for it by subtracting the expected bias from the entropy
standard deviation, given that the number of samples (IDS alarms)
is sufficiently large. \emph{However, this only makes sense if it
is known that the data are encrypted. }Since this in general is not
known for the payload from IDS rules, and it will be wrong to correct
for this bias for nonencrypted data, this means that the metric with
fastest convergence speed is preferable. 

It must also be noted that bit-based entropies (both Shannon Min-entropy)
are computationally less complex than octet-based Shannon entropy,
which needs to calculate the weighted logarithm expression for each
symbol in an octet. Counting the number of bits set to one in an octet
or word (list of octets) can be done by calculating the Hamming weight\index{Hamming weight},
which is implemented in hardware on most modern Intel or AMD processors
using the \emph{popcnt} (population count\index{population count})
operator. This opens up for efficient implementations of bit-entropy\index{bit-entropy}
calculations for up to 64 bits word chunks~\citep{intelSSE4reference},
which is more efficient than iterating to calculate the octet frequencies,
as required by octet-based entropies.

\subsection{\label{sub:Difference-Encrypted-Plaintext}Entropy Standard Deviation
Difference between Encrypted and Plaintext data}

Another foundational scenario that must be investigated, is how well
the proposed entropy algorithms and symbol definitions distinguish
between encrypted\index{encrypted} and plaintext\index{plaintext}
information. The entire theory behind $\sigma_{\alpha}$ hinges on
the assumption that there is a difference in entropy standard deviation
between plaintext and as a limit case encrypted information. To determine
whether this assumption is true or not, and which entropy configuration
that works best, we set up another Monte-Carlo simulation\index{Monte-Carlo simulation},
this time comparing the entropy standard deviation of plaintext data
with the entropy standard deviation of random uniform data for both
Min- and Shannon-entropy, using both bit and octet-based symbol definition. 

The experiment configuration calculates the average and the 95\% confidence
band\index{95% confidence band@95\% confidence band} ($\pm2\sigma)$
from an ensemble\index{ensemble} of 10000 experiments. Each experiment
calculates the standard deviation over 50 samples for varying input
data length in bits, assuming that this is the smallest number of
samples that in practice will be used to reliably distinguish between
encrypted and plaintext data. If less samples are used per experiment,
then the confidence band will widen out, meaning that longer payload
will be needed to reliably distinguish between encrypted and plaintext
data. There is in other words a tradeoff between the payload length and
the number of samples required to reliably detect encrypted content. 

Random uniform data was measured in a similar way as the previous
experiment. The plaintext data was extracted using randomly selected
contiguous quotes from the Brown corpus\index{Brown corpus} \citep{Brown_corpus},
with varying data length in bits along the x-axis. 

\begin{figure}
\begin{centering}
\includegraphics[scale=0.45]{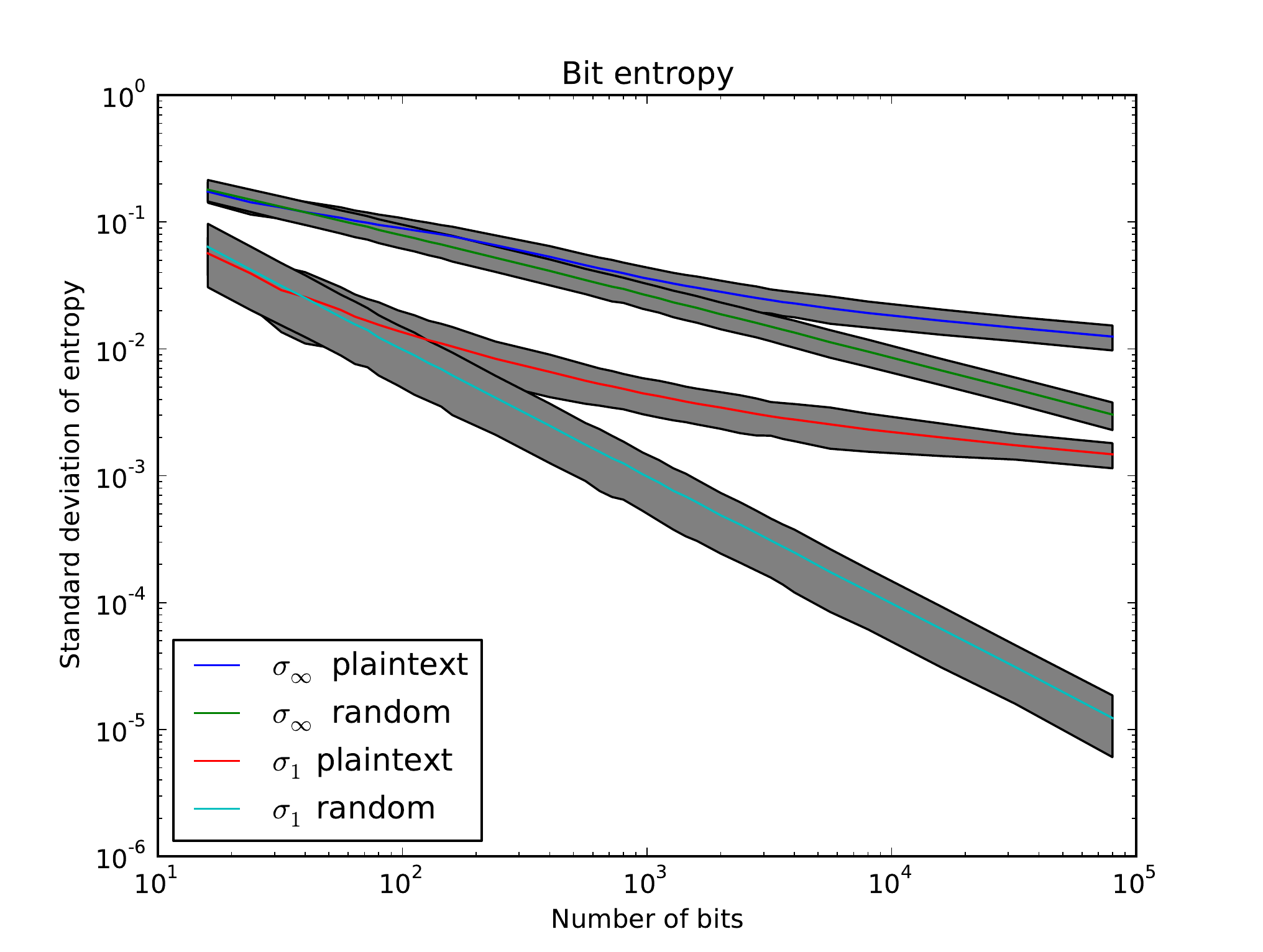}
\par\end{centering}

\caption{\label{fig:Difference-bit-entropy}Difference and 95\% confidence
band between $\sigma_{\alpha}$ for plaintext and random data using
bit-entropy for varying input data length in bits.}
\end{figure}

\begin{figure}
\begin{centering}
\includegraphics[scale=0.45]{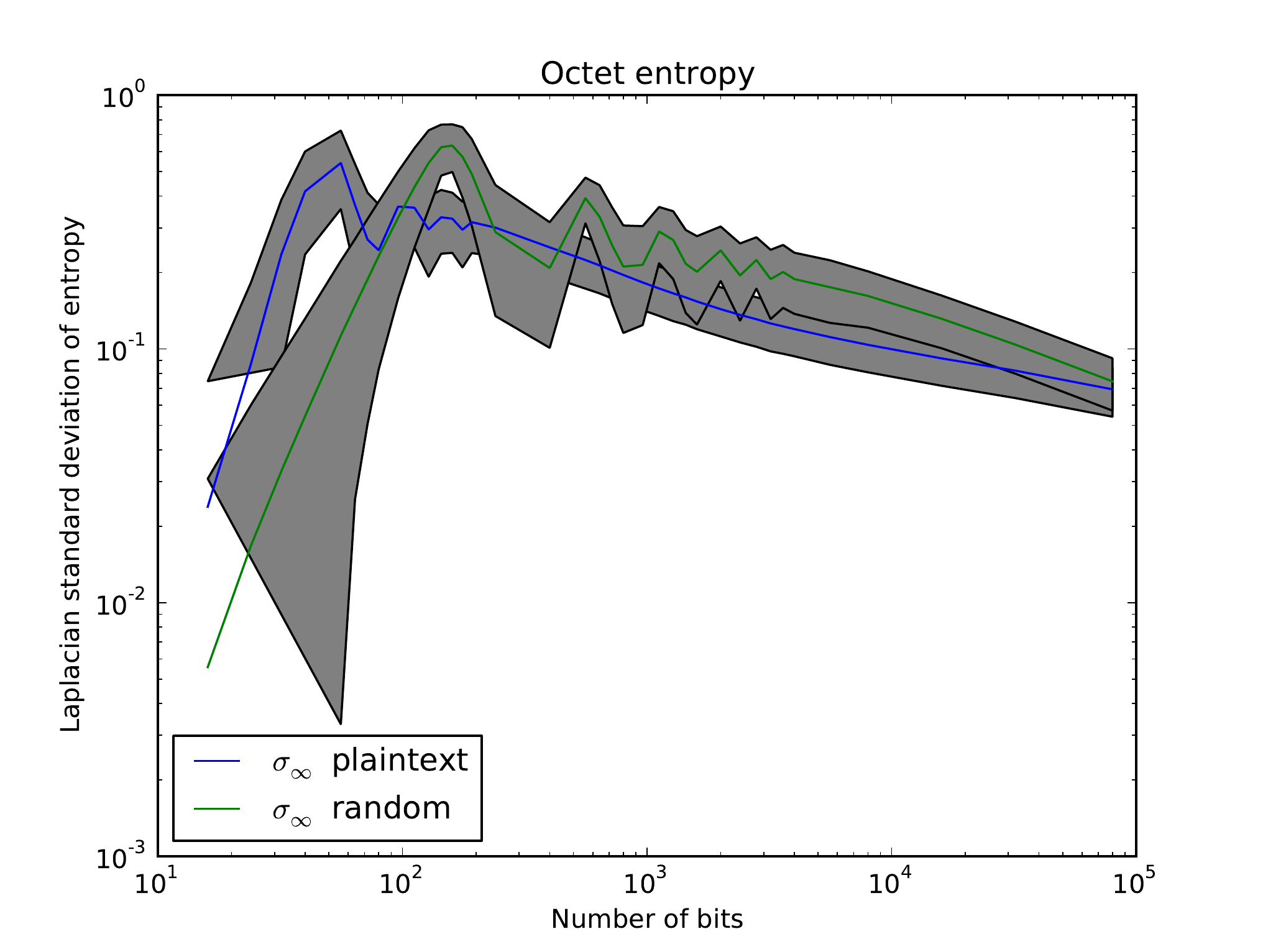}
\par\end{centering}

\caption{\label{fig:Difference-octet-min}Difference and 95\% confidence band
between $\sigma_{\infty}$ for plaintext and random data using Min-entropy
for varying input data length in bits.}
\end{figure}

\begin{figure}
\begin{centering}
\includegraphics[scale=0.45]{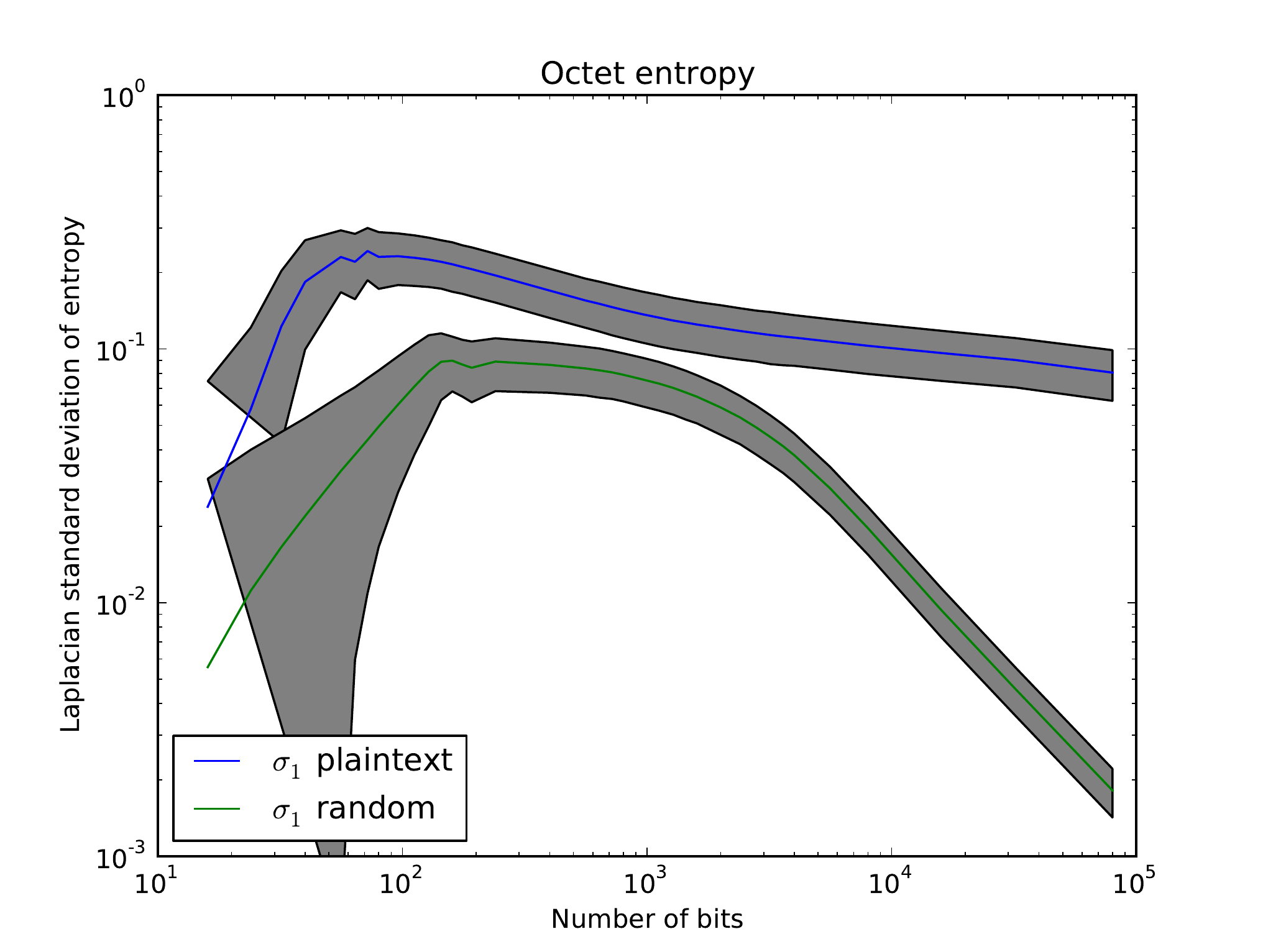}
\par\end{centering}

\caption{\label{fig:difference-normalised-shannon}Difference and 95\% confidence
band between $\sigma_{\alpha}$ for plaintext and random data using
normalised Shannon entropy for varying input data length in bits.}
\end{figure}

Figure \ref{fig:Difference-bit-entropy} shows the difference between
$\sigma_{\alpha}$ for plaintext and random data using bit- and octet-entropy
respectively for varying input data lengths in bits. Shannon bit-entropy\index{Shannon bit-entropy}
is the metric that distinguishes best between cleartext and encrypted
data for data lengths greater than 400 bits (50 bytes). 

Figure \ref{fig:difference-normalised-shannon} shows that Shannon
octet entropy is able to distinguish reliably between cleartext and
encrypted data over a sample of 50 IDS alarms within a 95\% confidence
interval from 5 octets (40 bits) and onwards, despite the poor convergence
properties for random traffic in the range $[5,131]$ octets.

However, due to the slightly hourglassed shape of the entropy difference,
it is not possible to achieve any larger precision between 40 and
3000 bits (375 bytes), unless the sample size is increased to narrow
the confidence band sufficiently. Plaintext data is 11 times larger
than encrypted data at 5 octets (40 bits), whereas at around 128 octets
(1024 bits), is down to 1.8 times larger than the encrypted data,
before the random data reaches its knee point where the octet-based
metric again improves. 

Shannon bit-entropy is more well-behaved than Shannon octet entropy\index{Shannon octet entropy},
in that the difference in entropy seems to be a strictly convex function,
as opposed to the octet-based entropies. Min-bit-entropy also seems
to be well behaved, and has the advantage that the 95\% confidence
band for Min-bit-entropy\index{Min-bit-entropy} is narrower than
for Shannon bit-entropy. However it is still overall a much poorer
measure of entropy difference than Shannon bit-entropy\index{Shannon bit-entropy},
since it requires at least 6000 bits (750 octets) to reliably distinguish
between plaintext and encrypted data. Octet-based Min-entropy, , as
shown in Figure~\ref{fig:Difference-octet-min}, behaves extremely
poorly, and is not usable for distinguishing between plaintext and
encrypted text. 

Overall, this strengthens the conclusion that Shannon entropy is the
best metric, regardless of symbol definition since it converges faster
than the other alternatives and it distinguishes better between cleartext
and encrypted data as long as the payload is longer than the minimum
threshold of 5 octets for octet-based entropy or 50 octets for bit-entropy
for minimum 50 samples.

\subsection{\label{sub:Payload-Length-Correction}Payload Length Correction for
Bit-entropy\index{payload length correction}}

\begin{figure}
\begin{centering}
\includegraphics[scale=0.45]{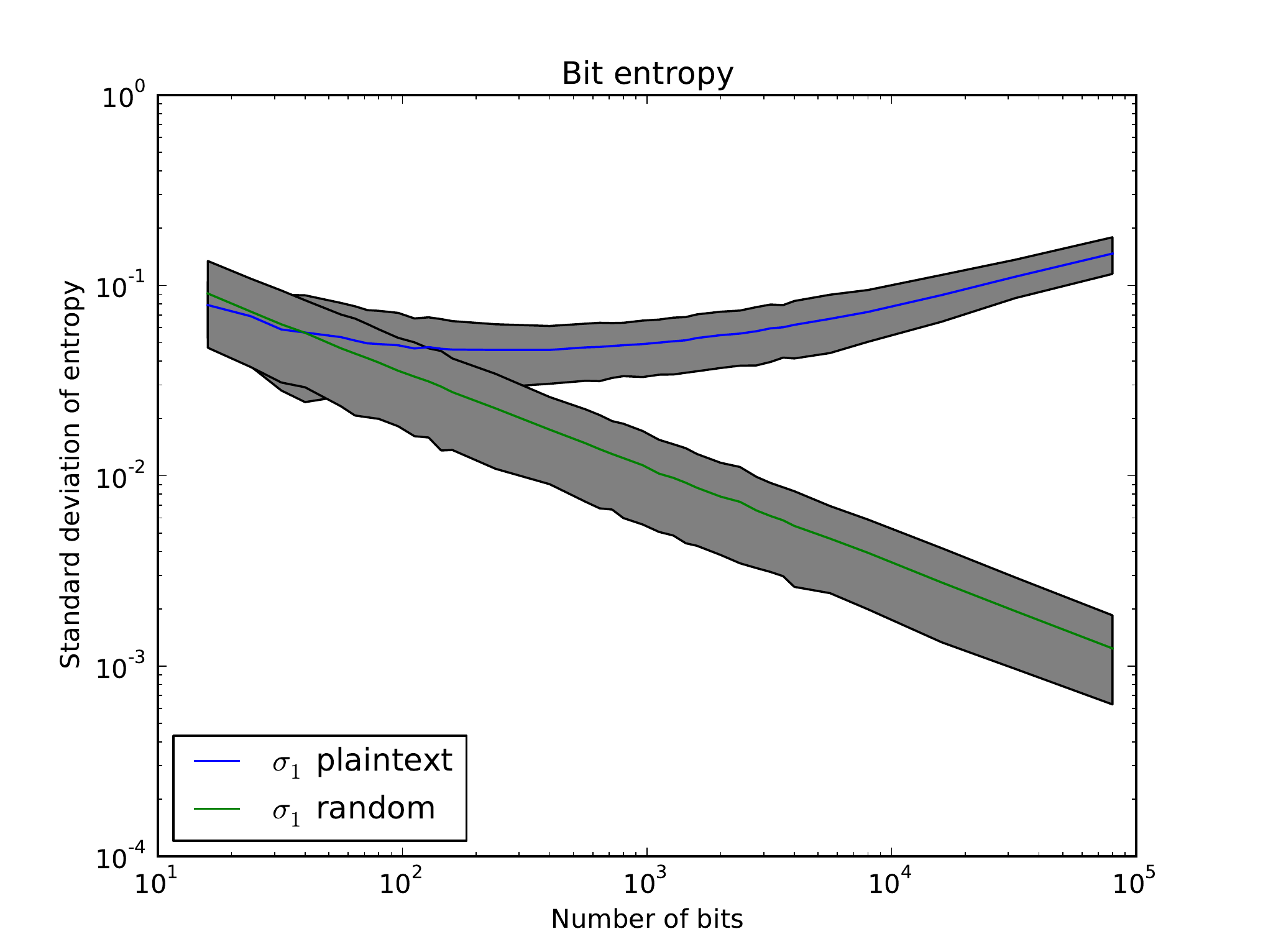}
\par\end{centering}

\caption{\label{fig:Payload-length-corrected}Payload length corrected Shannon
bit-entropy with 95\% confidence band as a function of input data
length in bits for plaintext and random data.}
\end{figure}

A deficiency with the entropy standard deviation metrics, is that
they decrease as the data length increases. This is the desired behaviour
for random uniform data, however it is not necessarily desirable for
plaintext data, since this means that the metric can not be considered
incentive compatible\index{incentive compatible}: it will then pay
off for an adversary to match as large plaintext data packets as possible,
since this in effect reduces the measured information leakage. An
obvious way to mitigate this problem might be to multiply the entropy
values with the length $n_{i}=\left|y_{i}\right|$ of the IDS alarm,
i.e. $n_{i}H_{\alpha}(y_{i})$, and then take the standard deviation
of the length corrected entropy values. This correction\index{correction}
will however be too strong, since the expected bias for random uniform
data of length $n_{i}$ then would be constant: $\sigma_{1}^{bias}\approx\frac{\gamma_{1}n_{i}}{n_{i}}=\gamma_{1}$.
This means that the metric would not converge to zero for encrypted
traffic.

This problem can be mitigated by multiplying the entropy values with
the \emph{square root} of the payload length $n_{i}$. This means
that the length corrected entropy\index{length corrected entropy}
values for bit-entropy can be described as $H_{\alpha}^{'}(y_{i})=\sqrt{n_{i}}H_{\alpha}(y_{i})$. 

The length-corrected privacy leakage metric $\pi_{R}^{L}$, can be
expressed as:

\begin{equation}
\pi_{R}^{L}=I\cdot\sigma_{k}^{L}=I\cdot\sqrt{2}\frac{{\displaystyle \sum_{i=1}^{N}w_{i,k}\left|H_{1}^{'}(y_{i})-\tilde{\mu}_{k}^{'}\right|}}{{\displaystyle \sum_{i=1}^{N}w_{i,k}}}.
\end{equation}

where $\tilde{\mu}_{k}^{'}$ is the median from the LMM.

\begin{figure}
\begin{centering}
\includegraphics[scale=0.45]{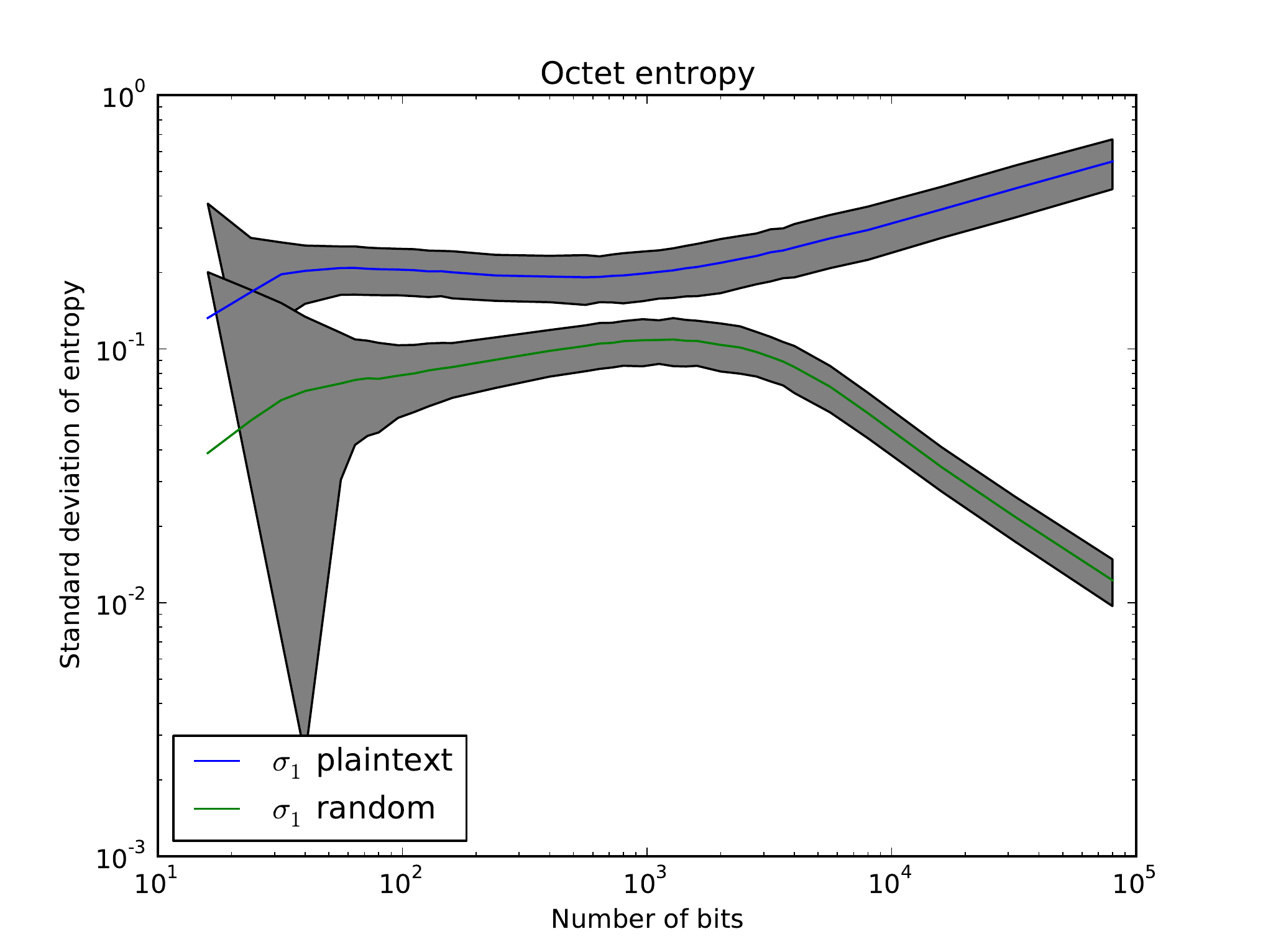}
\par\end{centering}

\caption{\label{fig:Payload-length-corrected-octet}Payload length corrected
Shannon octet-entropy with 95\% confidence band as a function of input
data length in bits for plaintext and random data.}
\end{figure}

The payload length corrected Shannon bit-entropy standard deviation
function is shown in Figure \ref{fig:Payload-length-corrected}. It
can be observed that the term $\sqrt{n_{i}}H_{\alpha}(y_{i})$ essentially
reduces the convergence speed to detect random uniform traffic for
Shannon entropy by a factor of $O(n^{-\frac{1}{2}})$ to $\sigma_{1}^{bias}\approx\frac{\gamma_{1}}{\sqrt{n}}$,
similar to Min-entropy originally. However random uniform traffic
will still converge towards 0, as required, although somewhat more
slowly. Furthermore, the measured privacy leakage for plaintext data
will now increase exponentially as a function of payload length, instead
of decreasing, as long as the payload length is larger than the required
100 bytes (800 bits). These modifications avoids the incentive incompatibility
for Shannon bit-entropy, since the metric now increases with increasing
payload length.

\subsection{Payload Length Correction for Shannon Octet-based Entropy}

Shannon octet-based entropy has the same convergence speed as Min-entropy
after an initial transient phase, as shown in Figure~\ref{fig:Log-log-plot-of-entropy-std}.
This means that $\sqrt{n_{i}}$ can be used as a length correction
factor also for Shannon octet-entropy for large entropy values (>200
octets or 1600 bits), to ensure that the measured privacy leakage
increases with the payload length for plaintext data, and decreases
with the payload length for random data.

This length correction does however not work well below 200 octets,
since Shannon octet entropy initially rises quicly until a knee point
at 50 bits for plaintext data and 150 bits for random data, and then
starts falling, as shown in Figure \ref{fig:difference-normalised-shannon}.
It is desirable to reduce the effect of this knee point, in order
to have an easier functional relationship between plaintext and random
data, so that a fixed threshold can be used to distinguish between
cleartext and random traffic. Introducing an additional length correction
factor of $\frac{1}{log_{2}(n_{i})}$ where $n_{i}$ is the length
of the payload $y_{i}$ can be used to reduce the effect of this knee
point, as shown in Figure \ref{fig:Payload-length-corrected-octet}.
This means that the payload length correction function for Shannon
octet-based entropy is $H_{\alpha}^{'}(y_{i})=\frac{\sqrt{n_{i}}}{log_{2}(n_{i})}$.

Payload length corrected Shannon octet-entropy standard deviation
as a function of payload length is shown in Figure \ref{fig:Payload-length-corrected-octet}.
The initial slightly hour-glassed shape of the standard deviation
functions means that the octet-based function despite the payload
correction still is reduced slightly for plaintext data between 48
and 800 bits (5 and 100 bytes) payload length. This means that the
metric is not entirely incentive compatible in this range, since it
is slightly decreasing for plaintext istead of increasing, however
the deviation is not very large. The octet-based metric is however
incentive compatible\index{incentive compatible} beyond 100 bytes,
since the metric then increases with increasing payload length for
plaintext data. An advantage with Shannon octet-entropy, is that it
is able to detect whether short strings of data is encrypted or cleartext,
for example from pseudonymisation schemes, assuming that the data
is at least 5 octets and encrypted using a perfect encryption scheme.

Another advantage with the payload length corrected entropy metrics,
is that a \emph{fixed threshold\index{fixed threshold}} can be used
to distinguish between plaintext and random data, regardless of payload
length for a sufficiently large sample (minimum 50 samples). For Shannon
bit-entropy this threshold is 0.028, whereas Shannon octet-entropy
has a threshold of 0.14 (five times larger).

It must be noted that it is possible to construct data that falls
between the two example entropies used here. The first example that
comes to mind, is partially encrypted IDS alarms, where for example
a header part is nonencrypted and a payload part is encrypted or coded
(e.g. compressed). In these cases, some IDS alarms would be interpreted
as encrypted, whereas others may be interpreted as nonencrypted. However,
an advantage with the octet-based metric, is that relatively few octets
are required to calculate it, which means that the header and remaining
payload in such cases can be calculated separately.

\begin{figure}
\begin{centering}
\includegraphics[scale=0.45]{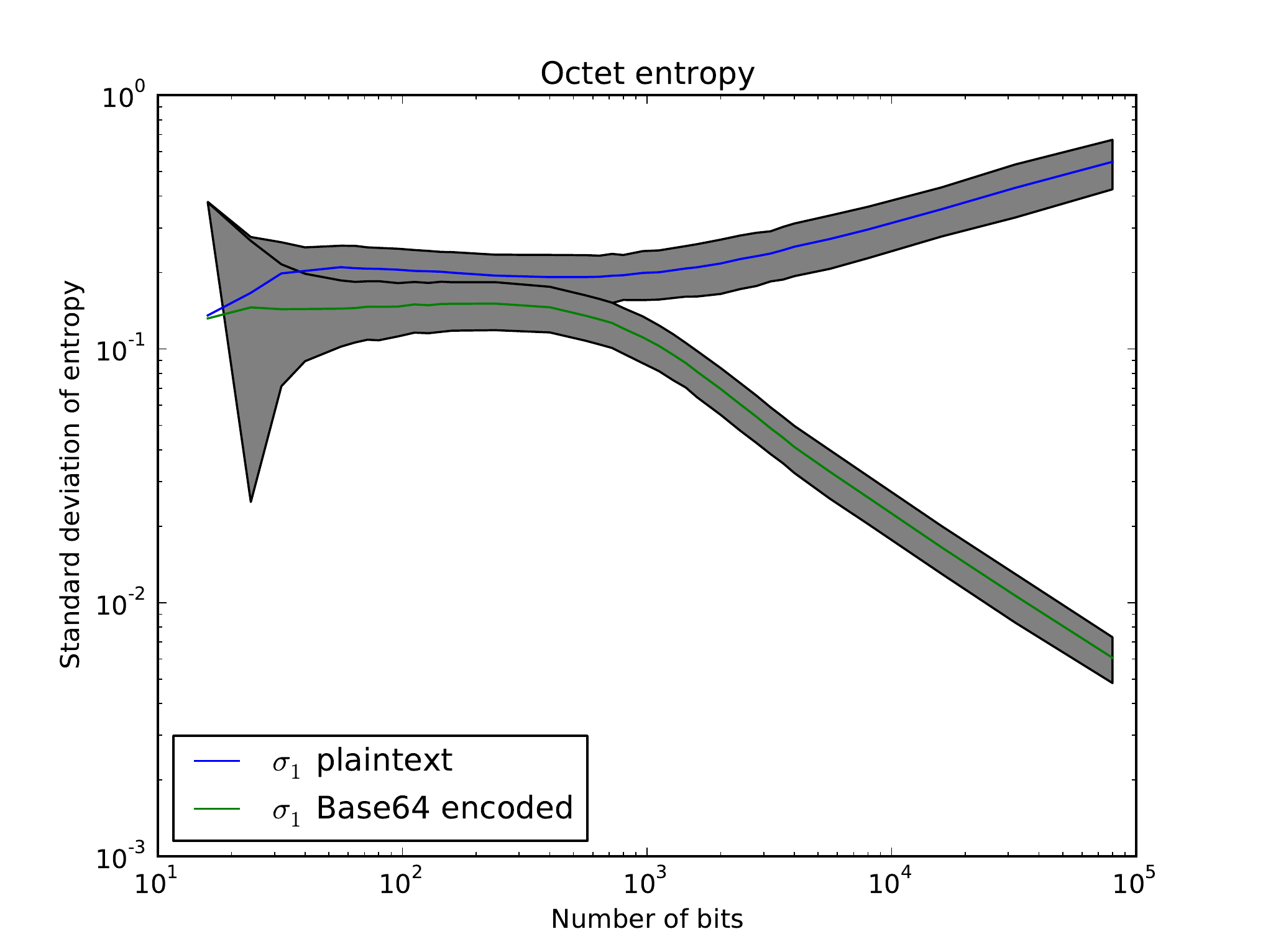}
\par\end{centering}

\caption{\label{fig:Payload-length-corrected-base64}Payload length corrected
Shannon octet-entropy with 95\% confidence band as a function of input
data length in bits for plaintext and Base64-encoded random data.}
\end{figure}

\subsection{Standard Deviation of Entropy for Base64-encoded Data}

Another interesting case is how $\sigma_{\alpha}$ copes with quoting
techniques used to transfer binary data on transport protocols that
are not 8-bit clean. A common encoding technique is Base64-encoding\index{Base64-encoding},
which can be used to transfer binary information in SMTP and XML-based
formats like HTML or SOAP. Figure \ref{fig:Payload-length-corrected-base64}
shows the standard deviation of Shannon octet-entropy for plaintext
and Base64 encoded random data. The Base64-encoding adds redundancy,
which means that the encoded data is closer to plaintext data. This
can be seen from the Figure, since the confidence bands now overlap
for less than 800 bit (100 bytes). However, for longer input data,
the Base64-encoded random data behaves in a similar way like plain
random data, since the standard deviation goes towards 0. 

This means that at least 100 bytes are required to reliably distinguish
between Base64-encoded random data and plaintext data. If it is known
that the information is Base64-encoded, then it will be possible to
decode the information before the entropy is calculated. This may
be useful if the information leakage of shorter Base64-encoded strings
are being measured. However this decoding will add additional parsing
overhead, which may not be desirable from a performance perspective.
This is however avoidable, as long as the payload is larger than 100
bytes as shown above.

\subsection{Semantic Information of Symbols\index{semantic information of symbols}}

The symbol definition for the entropy algorithms will also need to
take into account the semantic information that symbols convey. The
definition of bytes (or more precisely octets) is in particular important
for computer systems, since this is used to define the basic character
set used for communicating both text and binary codes. Octet-based
symbol definition is also important for many of the attack vectors
discussed in the introduction. Buffer overflow attacks for example
frequently use the single octet NOP instruction (0x90 on Intel machines)
for the NOP sled\index{NOP sled}. There also exist multi-octet NOP
variants and other techniques for generating an obfuscated sled \citep{akritidis_stride:_2005}.
However for now consider single byte based NOP sleds, which are common,
not the least because they are easier to exploit. Using this strategy
means that the shellcode does not need to be placed on an exact 32-
or 64 bits word boundary, as compilers typically enforce for normal
programs \citep{akritidis_stride:_2005}.

The single-byte NOP sled (0x90) is a unique symbol for octet-based
entropy\index{octet-based entropy}, however for bit entropy, this
represents the binary string $10010000$, which has two out of eight
bit set. The problem is that this value is not unique. There will
in general be $8!/((8-2)!\cdot2!)=28$ different octets, where any
combination of these can produce the same two-bit based entropy value
as this NOP opcode. In fact, bit-entropy means that 256 different
octet values are mapped down to only 9 different bit-entropies. Furthermore,
whereas the octet entropy of a list of NOP opcodes will be zero, the
bit-entropy will be greater than zero, except if all bits are '1'
or '0'. The Shannon bit-entropy of the NOP sled is 0.81, which is
very different from the octet-entropy (0). Furthermore, if one octet
of information is changed, this means that somewhere between one and
eight bits will change. There is in other words a less clear correlation
between the change in information and change in entropy for bit-entropy
than for octet-based entropy. 

This means that octet-based entropy is closer to representing the
\emph{meaning\index{meaning}} of the information being exchanged,
and therefore should be the preferred symbol definition for the privacy
metrics. The discussion above has also identified that the standard
deviation of Shannon octet-entropy is the metric that overall has
the best properties for distinguishing between cleartext and encrypted
data, despite its poor convergence properties over part of the usable
range. Octet-based entropy is furthermore able to uniquely identify
that a sequence of the same octet has zero entropy\index{zero entropy},
something bit-entropy does not identify. This means that Shannon octet-entropy
will provide the largest possible difference in entropy between plaintext
and strings consisting of sequences of the same character. Shannon
octet-entropy is in other words a better privacy leakage metric than
Shannon bit-entropy with better distinguishing capability according
to our requirements and needs within the operating range. Min-entropy
is not usable for our purpose.

\emph{}

\section{\label{sub:Analysis-of-Alarm-PDF}Experimental Results\index{experimental results}}

The experimental results are based on IDS alarms from my own home
network between 2009 and 2011. Some of the IDS alarms are also from
the KDD-Cup'99\index{KDD-Cup'99}\index{1999 KDD Cup data set} data
set. We included the 32 most noisy IDS rules with at least 50 IDS
alarms per cluster in the measurements. The threshold of 50 IDS alarms
per cluster is chosen to stay within the 95\% confidence band\index{95% confidence band@95\% confidence band}s
discussed in the simulations in Section \ref{sub:Difference-Encrypted-Plaintext}.
This is a limited data set that will not reflect the privacy leakage
measured at a professional MSS provider doing large-scale measurements.
The main difference that can be expected from a larger MSS provider,
is that there would be a greater selection of IDS alarms with more
than 50 alarms per cluster, and that the number of attack clusters
would be greater. Furthermore, a larger set of IDS alarms may be enabled
by commercial MSS providers to counter for emerging threats that are
not yet in the Snort VRT ruleset, which we used. Furthermore, traffic
from a commercial MSS provider would not be influenced by the synthetic
KDD-Cup'99 data set. 

However, despite these deficiencies, there are also some advantages
by using our own data. One of the main advantages, is that this allows
for discussing the IDS rules that may be leaking private or confidential
information in detail, something that it according to our experience
would be difficult or impossible to do for a commercial MSS provider
due to business confidentiality and repudiation concerns. We have
attempted to get agreement for such measurements for commercial MSS
providers, however this is only possible if the IDS ruleset is not
revealed, which makes it difficult to discuss in a convincing way
that the proposed privacy leakage metrics work as intended. More elaborate
tests at a commercial MSS provider is therefore left as future work.
We decided to use a privacy impact factor\index{privacy impact} $I=1$
to only show the information leakage part of the privacy leakage metrics.

The experiment includes an IDS rule that we created \emph{(sid:1:1394000)}
which tests the worst-case scenario\index{worst-case scenario} from
a privacy perspective. This is a threshold-based IDS rule that essentially
samples every 10th packet from the network. This is intended to show
the maximum value that the privacy metric typically is able to detect,
which is useful to see how far away the IDS rules in the measurements
are from a worst-case scenario.

\subsection{Number of Attack Vectors}

\begin{figure}
\begin{centering}
\includegraphics[scale=0.65]{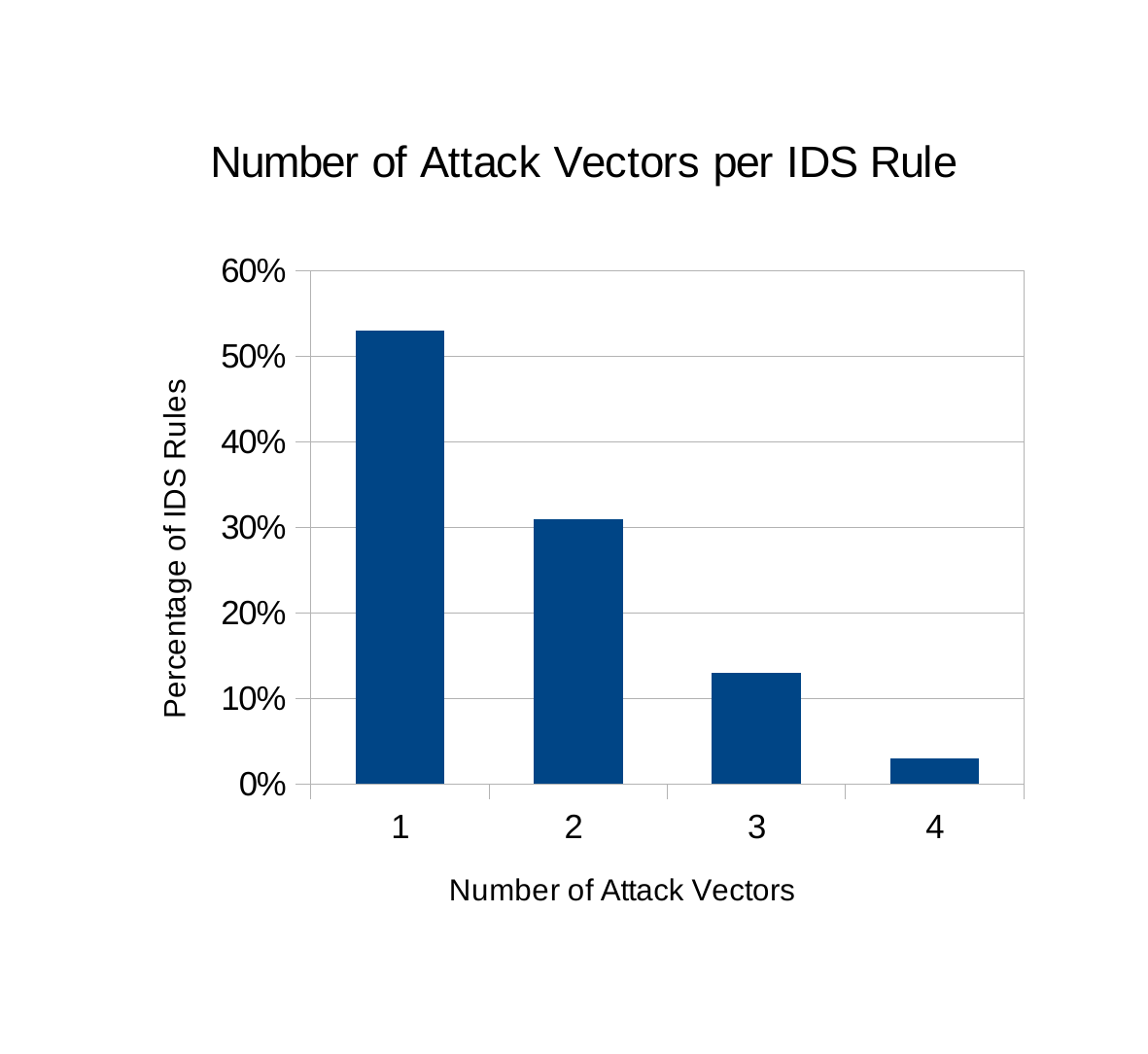}
\par\end{centering}

\caption{\label{fig:Number-of-attack-vectors}Number of attack vectors estimated
per IDS rule for Shannon octet-entropy.}
\end{figure}

The number of attack vectors\index{number of attack vectors} per
IDS rule for the given experiment is summarised in Figure \ref{fig:Number-of-attack-vectors}.
For this experiment, 53\% (17 rules) have one attack vector, 31\%
(10 rules) have two clusters identifying attack vectors, 13\% (4 rules)
have three clusters and 3\% (1 rule) have 4 clusters identifying attack
vectors. Please note that these numbers are specific to the given
experiment. A preliminary experiment at a commercial MSS provider
indicates that large-scale operations can expect the distribution
to be shifted somewhat towards more attack vectors. It is in other
words common that IDS rules may trigger on more than one attack vector,
which means that clustering must be used to calculate the entropy
of each underlying attack vector.

\section{Influence by Outliers\index{outliers}}

Figure \ref{fig:Standard-deviation-of-all} shows the Normal standard
deviation\index{Normal standard deviation} $\sigma_{1}$ and Laplacian
standard deviation\index{Laplacian standard deviation} $\sigma_{1}^{L}$
based on the $L^{1}$ norm for length corrected normalised Shannon
octet-entropy. The Figure shows that the Normal standard deviation
$\sigma_{1}$ for some IDS rules indicate a significantly larger privacy
leakage than the Laplacian standard deviation $\sigma_{1}^{L}$. The
most extreme cases are SID 119:14 which detects non-standard characters
in web requests and SID 1:399 ICMP Host unreachable. The reason for
the deviation is in both these cases outliers far out from the main
cluster. The Normal standard deviation will give too high weight to
the outliers in these cases, since it measures the root of the squared
distances. Other IDS rules where the Normal standard deviation of
entropy is somewhat influenced by outliers are amongst others SIDs
119:4, 119:15 and 1:1201.

In all these cases, the Laplacian standard deviation will give a more
realistic estimate of the privacy leakage than the Normal standard
deviation. The Laplacian standard deviation is only significantly
larger than the Normal standard deviation for SID 1:402 ICMP Destination
Port unreachable. This IDS rule has a left skewed noisy distribution,
with several peaks reflecting the servers that were attempted contacted,
but did not respond. We interpreted this as one cluster, since the
failed services strictly speaking cannot be considered attack vectors.
The median for this IDS rule (at 7.5) deviates somewhat from the mean
(at 7.2), which gives more weight to the leftmost peaks for the Laplacian
standard deviation than the Normal standard deviation does in this
case, causing the Laplacian standard deviation to be larger than the
Normal standard deviation. This is a pathological case\index{pathological case}
where the normal standard deviation may give a better estimate than
the Laplacian standard deviation. However, overall the Laplacian standard
deviation $\sigma_{1}^{L}$ should be used to calculate the privacy
leakage metric, since this in most cases is the more robust statistic\index{robust statistic}.

\subsection{Measured Information Leakage\index{measured information leakage}}

\begin{figure*}
\begin{centering}
\includegraphics[bb=50bp 70bp 842bp 454bp,clip,scale=0.54]{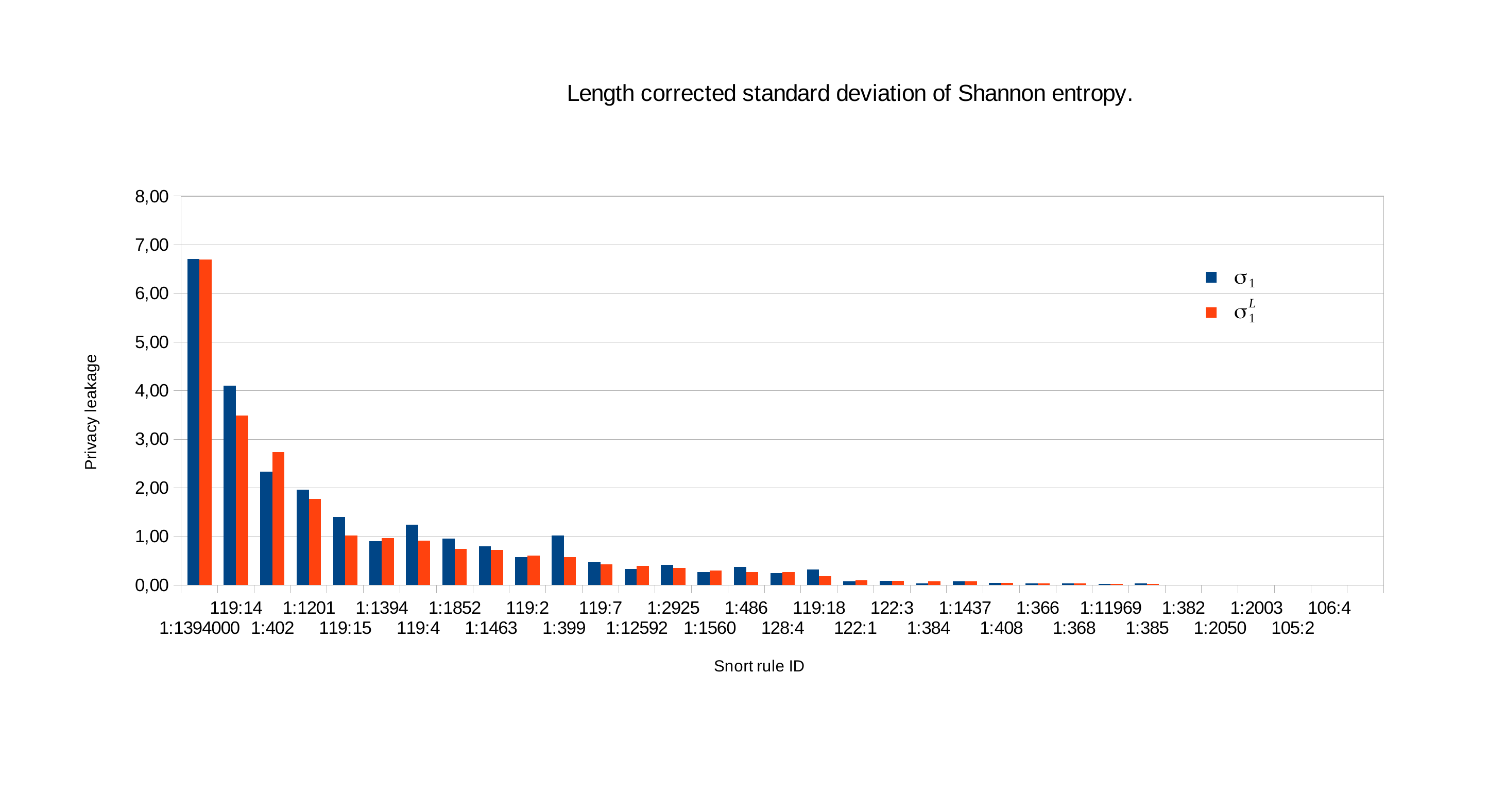}
\par\end{centering}

\caption{\label{fig:Standard-deviation-of-all}Privacy leakage measured using
length corrected standard deviation (Normal $\sigma_{1}$ and Laplacian
$\sigma_{1}^{L}$) of normalised Shannon octet-entropy as a function
of Snort IDS rule.}
\end{figure*}

\begin{table*}
\begin{centering}
\begin{tabular}{|l|r|c|c|c|>{\raggedright}p{7cm}|}
\hline 
Snort SID & Alarms & Clusters & $\sigma_{1}$ & $\sigma_{1}^{L}$ & Description\tabularnewline
\hline 
\hline 
1:1394000 & 95096 & 1 & 6,71 & 6,70 & Samples random traffic\tabularnewline
\hline 
119:14 & 3104 & 1 & 4,10 & 3,49 & http\_inspect non-standard characters in web request\tabularnewline
\hline 
1:402 & 36224 & 1 & 2,34 & 2,73 & ICMP Destination Port unreachable\tabularnewline
\hline 
1:1201 & 680 & 1 & 1,96 & 1,77 & HTTP 403 Forbidden\tabularnewline
\hline 
119:15 & 720 & 1 & 1,40 & 1,02 & http\_inspect over-long URL\tabularnewline
\hline 
1:1394 & 1384 & 2 & 0,90 & 0,97 & Shellcode x86 NOP AAAAAA\tabularnewline
\hline 
119:4 & 576 & 1 & 1,24 & 0,91 & http\_inspect preprocessor (IIS decoding attacks)\tabularnewline
\hline 
1:1852 & 10392 & 1 & 0,96 & 0,75 & robots.txt access\tabularnewline
\hline 
1:1463 & 288 & 1 & 0,80 & 0,72 & IRC Chat\tabularnewline
\hline 
119:2 & 21744 & 2 & 0,58 & 0,61 & http\_inspect double encoded characters\tabularnewline
\hline 
1:399 & 631840 & 1 & 1,02 & 0,58 & ICMP Host unreachable\tabularnewline
\hline 
119:7 & 1520 & 2 & 0,48 & 0,43 & http\_inspect unicode encoded web request\tabularnewline
\hline 
1:12592 & 312 & 1 & 0,33 & 0,40 & SMTP command injection attempt\tabularnewline
\hline 
1:2925 & 12960 & 2 & 0,42 & 0,35 & 1x1 GIF attempt (web bug)\tabularnewline
\hline 
1:1560 & 360 & 2 & 0,27 & 0,30 & WEB-MISC /doc access\tabularnewline
\hline 
1:486 & 368 & 1 & 0,37 & 0,27 & ICMP Destination Unreachable\tabularnewline
\hline 
128:4 & 306616 & 3 & 0,25 & 0,27 & spp\_ssh\tabularnewline
\hline 
119:18 & 22760 & 2 & 0,32 & 0,18 & http\_inspect directory traversal outside web server root.\tabularnewline
\hline 
122:1 & 576 & 2 & 0,08 & 0,10 & sfPortscan preprocessor (tcp portsweep)\tabularnewline
\hline 
122:3 & 2088 & 1 & 0,09 & 0,09 & sfPortscan preprocessor (tcp portsweep)\tabularnewline
\hline 
1:384 & 566016 & 4 & 0,04 & 0,08 & ICMP Ping (general)\tabularnewline
\hline 
1:1437 & 1056 & 2 & 0,08 & 0,08 & MULTIMEDIA Windows Media download\tabularnewline
\hline 
1:408 & 205904 & 3 & 0,04 & 0,04 & ICMP Echo Reply\tabularnewline
\hline 
1:366 & 202552 & 1 & 0,04 & 0,04 & ICMP Ping {*}NIX\tabularnewline
\hline 
1:368 & 202552 & 1 & 0,04 & 0,04 & ICMP Ping BSD\tabularnewline
\hline 
1:11969 & 2896 & 3 & 0,03 & 0,03 & VOIP-SIP inbound 401 Unauthorized\tabularnewline
\hline 
1:385 & 4392 & 2 & 0,04 & 0,03 & ICMP traceroute\tabularnewline
\hline 
1:382 & 2192 & 1 & 0,00 & 0,00 & ICMP Ping Windows (alphabet)\tabularnewline
\hline 
1:2050 & 32024 & 1 & 0,00 & 0,00 & SQL Version Overflow attempt.\tabularnewline
\hline 
1:2003 & 1777264 & 1 & 0,00 & 0,00 & SQL Worm Propagation attempt.\tabularnewline
\hline 
105:2 & 192 & 2 & 0,00 & 0,00 & BO traffic (spp\_bo)\tabularnewline
\hline 
106:4 & 464 & 3 & 0,00 & 0,00 & spp\_rpc\_decode preprocessor - e.g. incomplete RPC segment.\tabularnewline
\hline 
\end{tabular}
\par\end{centering}

\caption{\label{tab:Measured-Privacy-leakage}Privacy leakage measured using
length corrected standard deviation based on Shannon octet-entropy
for the IDS rules in the experiment.}
\end{table*}

Figure \ref{fig:Standard-deviation-of-all} shows the measured privacy
leakage for the experiment using length corrected standard deviation
(Normal $\sigma_{1}$ and Laplacian $\sigma_{1}^{L}$) of normalised
Shannon octet-entropy as a function of Snort IDS rule. Further details
can be found in Table \ref{tab:Measured-Privacy-leakage}. This discussion
is based on the Laplacian standard deviation\index{Laplacian standard deviation},
since the previous section shows that the Normal standard deviation
has problems with outliers in the dataset. First, it can be observed
that the metric works as expected for the extreme cases. The IDS rule
that performs random sampling of payload (SID 1:1394000) has the highest
privacy leakage\index{highest privacy leakage}. On the other hand,
there also exist 5 IDS rules that are very precise at matching the
attack vector, and behaves like the perfect model IDS rule $R_{P}$
with zero privacy leakage. IDS rules that fall into this category
are attack vectors like SID 1:2050 SQL Version Overflow attempt, SID
1:2003 SQL Worm Propagation attempt, SID 105:2 BO traffic and SID
106:4 spp\_rpc\_decode preprocessor which detect amongst others incomplete
RPC segments. All these IDS rules indicate possibly malicious activities\index{malicious activities},
and are precise at detecting the attack. SID 1:382 which detects ICMP
Echo requests (Ping) for Windows also behaves like a perfect IDS rule.
It typically sends the alphabet in the payload. 

There are furthermore 9 additional IDS rules with privacy leakage
lower than the threshold of 0.14 for distinguishing between plaintext
and encrypted traffic that was identified in Section~\ref{sub:Payload-Length-Correction}.
Rules in this category can be considered to have insignificant privacy
leakage, since it is not distinguishable from encrypted traffic. These
include ICMP rules matching ICMP Echo Request and Reply for various
platforms (SIDs 1:384, 1:408 and 1:368) and ICMP traceroute (SID 1:385).
These ICMP protocols are part of the TCP/IP protocol suite and are
benign in themselves, however the Ping protocol is also frequently
used for malicious activities like Denial of Service attacks or Ping
scans. Furthermore pre-attack activities like portscanning (SIDs 122:1
and 122:3), and unauthorised inbound SIP calls (SID 1:11969) are potentially
malicious activities that fall into this category. Last, SID 1:1437
detects download of Windows media files. This would normally be considered
a benign activity, and it may also be concerning from a privacy perspective
if this IDS rule is activated, since it could be used to monitor user
activities. This rule detects download of Windows media files as two
narrow clusters, where the upper cluster at an entropy close to 1
probably indicates download of the compressed media file. This is
an example of a pathological case where the entropy standard deviation
in itself, as an indirect measure of privacy leakage, does not match
the perceived privacy leakage. The data controller may in this case
consider whether the privacy impact $I$ of this IDS rule should be
increased.

The privacy leaking IDS rules can broadly be subdivided into two groups:
IDS rules with large privacy leakage\index{large privacy leakage}
($\sigma_{1}^{L}>1)$ and IDS rules with medium privacy leakage ($\sigma_{1}^{L}\in<0.14,1]$).
There are 13 IDS rules with medium privacy leakage\index{medium privacy leakage}.
The most privacy leaking of these IDS rules, is SID 1:1394 ``SHELLCODE
x86 inc ecx NOP'' which triggers on any packet that contains a sequence
of 31 'A' characters ($\sigma_{1}^{L}=0.97$). The problem is that
this sometimes occurs in hex-encoded URLs or hex-encoded data in web
pages. It may also occur in non-compressed images, as well as for
other protocols. This means that the rule most likely will trigger
on a lot of random traffic, which is problematic from a privacy perspective. 

Many of the rules with medium privacy leakage may be triggered by
normal user behaviour, for example SID 1:486 ICMP Destination Unreachable,
SID 1:402 ICMP Destination Port Unreachable and SID 1:399 ICMP Host
unreachable. These can be problematic from a privacy perspective,
since the ICMP error message often contains the payload of the original
request, and these error messages can for example be triggered by
high traffic volume (or DoS attacks) towards a server. This means
that these ICMP messages essentially sample random user traffic\index{sample random user traffic}.
There are also other IDS rules in this category that will sample random
traffic from users, which for example may be used in user profiling.
Examples of such rules are SID 1:2925 1x1 GIF attempt that detects
web bug\index{web bug}s, SID 1:1560 that triggers on access to /doc
on the web server root and SIDs 119:2, 119:4, and 119:7 that aim at
detecting anomalies in HTTP requests like double encoded requests,
IIS decoding attacks\index{decoding attacks} and unicode encoded
requests. These may indicate attacks, but will in most cases probably
be false alarms that essentially sample random user traffic, something
that may be problematic from a privacy perspective. 1:1852 robots.txt
access, which normally indicates indexing of a web server by a web
crawler, also falls into this category. 

There are also some other attack rules with medium privacy leakage
that do not target ICMP or web traffic. SID 128:4 detects non-SSH
traffic on an SSH port, or a protocol mismatch\index{protocol mismatch}
(e.g. SSH1 traffic on an SSH2 port). This rule triggers on the initial
key negotiation phase, where some information in the SSH protocol
goes in cleartext. This can probably not be considered a significant
primary source of privacy leakage, since no sensitive information
is transferred in the packets. The data controller may consider reducing
the privacy impact for this IDS rule. SID 1:12592 detects SMTP command
injection attempts, that aims at exploiting a bug in the ClamAV anti-virus\index{anti-virus}
system. The rule definition is a very simple regular expression which
is likely to have false alarms. This rule may therefore be concerning
from a privacy perspective, although it mostly triggered on spam.
Rule 1:1463 triggers on IRC chat traffic, which also may be concerning
from a privacy perspective. The reason for implementing this rule,
is that IRC bots\index{IRC bots} also often have been used to control
botnets\index{botnets} of compromised hosts\index{compromised hosts}.
However, the rule does not check whether the traffic is benign or
not.

There are four IDS rules with high privacy leakage, not including
the test rule that samples random traffic. Three of these trigger
on web traffic: SIDs 119:14, 119:15 and 1:1201. The most privacy leaking
ordinary IDS rule (SID 119:14, $\sigma_{1}^{L}=3.49$) triggers on
non-standard character encodings in HTTP requests, which are getting
increasingly common, especially after IANA allowed non-ASCII domain
names. The second most privacy leaking IDS rule is SID 1:402 ICMP
Destination Port Unreachable with $\sigma_{1}^{L}=2.73$. This protocol
typically copies the failed request in the ICMP message, and therefore
samples random traffic requests. On third place is SID 1:1201 HTTP
403 Forbidden ($\sigma_{1}^{L}=1.77$), which also is quite common
also for benign traffic, for example on web sites referring to internal
material that require subscription. On fourth place is SID 119:15
that tests for over-long URL's ($\sigma_{1}^{L}=1.02$), something
that frequently happens for blogs or search engines that use URL referencing.
All of these rules may be problematic from a privacy perspective,
since they in many cases will trigger on normal user behaviour. It
is especially problematic if the IDS rules monitoring web services
are set up in an uncritical way, so that these rules trigger for any
web server accesses and not only for relevant web servers (e.g. the
company's own web servers).

This discussion shows that the privacy leakage metric\index{privacy leakage metric}
is able to distinguish between IDS rules that most likely may trigger
on ordinary user activities, and therefore may be problematic from
a privacy perspective, from the IDS rules that are precise at detecting
the underlying attack vector, or that perform a very specific task
without leaking any significant amount of data about user behaviour.
However there were also two pathological case\index{pathological case}s
where it may make sense to adjust the privacy impact\index{privacy impact},
since using entropy as an indirect measure of privacy leakage not
always will gives a true picture of whether the underlying information
is sensitive from a privacy/confidentiality perspective or not. Overall,
this demonstrates that the privacy leakage metric works as intended.
However larger studies involving commercial MSS providers will be
needed in the future to confirm these results.

\subsection{The Effect of Anonymisation\index{effect of anonymisation}}

The resulting privacy leakage over all IDS alarms in the experiment,
weighted according to number of alarms, is 0.31. However, if the test
IDS rule with SID 1:1394000 that samples random data is removed, then
the resulting privacy leakage is reduced to 0.16. If all the IDS rules
with high privacy leakage are removed, then the resulting leakage
is reduced by 0.02 to 0.14. 

Surprisingly, it is then more efficient to anonymise all ICMP Destination
Host unreachable alarms, since there are many of them (631840) in
the data set, and each of them has a significant measured privacy
leakage ($\sigma_{1}^{L}=0.58$). Anonymising ICMP Destination Host
unreachable alarms would reduce the overall privacy leakage by 0.07
to 0.07. This can probably be done without reducing the usability
for the security analyst significantly, since it still would be known
which host that was attempted contacted from the IP-address element
of the IDS alarm. SID 1:402 ICMP Destination Port unreachable also
triggers quite often (32360 times) and has the second highest measured
privacy leakage ($\sigma_{1}^{L}=2.73$). Anonymising this rule reduces
the privacy leakage by 0.02 to 0.05, and can probably also be done
without reducing the possibility to do root cause analysis significantly,
since the number of services running on a server normally is limited.
Classification based on the EM-clustering can if necessary be used
to indicate which server that failed without revealing the original
user request. These examples show that the total privacy leakage,
calculated as the product of number of IDS alarms $N_{R}$ for the
given rule $R$ and the entropy standard deviation $\sigma_{1,R}^{L}$,
must be used as the optimisation criterion to reduce the overall privacy
leakage. The total privacy leakage is calculated as $L_{tot}=N_{R}\sigma_{1,R}^{L}$. 

Another IDS rule, that either benefits from anonymisation, alternatively
by setting the privacy impact to zero, is SID 128:4 which detects
ssh anomalies. This rule triggers quite often (306616 times) with
$\sigma_{1}^{L}=0.27$, which means that the overall privacy leakage
can be reduced by 0.02 to 0.03 if this rule is anonymised. If the
IDS rules with low privacy leakage, that are not relevant from a privacy
perspective (all with privacy leakage less than 0.14, except SID 1:1437),
are either anonymised or removed by setting the privacy impact to
zero, then the resulting privacy leakage index is reduced from 0.03
to 0.011. 

If the two IDS rules from the http\_inspect preprocessor with largest
total leakage (SID 119:14 and 119:2) also are anonymised, then the
measured privacy leakage is reduced to 0.005.

This illustrates how a structured method can be used to reduce the
privacy leakage\index{reduce the privacy leakage} of the IDS ruleset
based on measured privacy leakage and number of IDS alarms. It is
furthermore also clear that many of the IDS rules can be anonymised
without significantly reducing the usability\index{usability} for
the security analysts\index{security analyst}. Especially since the
clustering model used to identify attack vectors in many cases can
be used to help the security analysts in identifying the necessary
properties of the underlying data without having to reveal the user
payload.

\emph{}

\section{\label{sec:Related-Works}Related Works}

\emph{}

The research area of quantitative information flow based on information
theory adds a comprehensive theoretical framework for analysing privacy
leakage based on entropies~\citep{smith_foundations_2009,smith_quantifying_2011}\@.
Our research is based on this, and extends the theory to cover privacy
leakage in IDS alarms. There is as far as we are aware of no other
research that proposes a comprehensive model of privacy leakage in
IDS alarms based on quantitative information flow analysis. 

Quantitative information flow analysis that in a similar way uses
information entropy has however been proposed used to derive an intrusion
detection capability metric in \citep{gu_measuring_2006}. This metric
aims at modelling the uncertainty about the input given the IDS output.
The uncertainty as it is termed in this paper is the same as the information
leakage defined here based on \citep{smith_foundations_2009}, which
in turn is based on the notion of mutual information from \citep{shannon_1948}.
The IDS capability metric is defined as the mutual information between
the IDS input and output to the entropy of the input:
\begin{equation}
C_{ID}=\frac{H(X)-H(X|Y)}{H(X)}
\end{equation}

The numerator is the same as the information leakage defined in \citep{smith_foundations_2009},
however these data are normalised with respect to the entropy of the
input data, something our model does not do. This model assumes that
the input data $H(X)$ is the labels (attack or not) from a labeled
IDS test set, and the output data $H(X|Y)$ is the classification
by the IDS, which also is different from our conceptual model of an
IDS rule. It is from this clear that the proposed metric is different
from the privacy leakage metric proposed here, since it assumes different
input data, a different information model and normalises the indicator
to the input data. However an interesting similarity is that the effect
of false positives in Figure 3b) in this paper follows a similar falling
exponential curve as Figure \ref{fig:Standard-deviation-of-all},
as can be expected, since the false alarms here will increase the
entropy up to the point where the classifier is not better than random
decisions. However this paper does not make the connection to privacy
leakage metrics for IDS rules.

There are also some similarities between the proposed approach and
the concept of Differential Privacy in statistical databases~\citep{dwork_differential_2006,dwork_differential_2008,dwork_differential_2009}.
Both methods use a Maximum Likelihood (ML) estimate, however the estimate
is interpreted differently. Differential Privacy uses the ML estimate
to indicate the aggregate value of underlying perturbed data, whereas
we use the ML estimate as a measure of underlying attack vectors.
Both methods use robust statistics (first norm) for calculating aggregated
values. However, Differential Privacy typically adds Laplacian noise
to hide individual elements of privacy sensitive information, whereas
our privacy leakage metric works in the opposite way - assumed Laplacian
noise from an IDS rule is used as an indication of IDS privacy leakage.
So although there are similarities, our proposed metric is clearly
different to Differential Privacy.

Entropy has previously been proposed as a measure of privacy \citep{clau_structuring_2006,bezzi_entropy_2008}.
Claude Shannon's seminal paper on information theory was the first
publication where entropy was proposed to measure the level of ambiguity
or equivocation in transferred information~\citep{shannon_1948}.
Min-entropy has been proposed as a metric of anonymity that in particular
considers local aspects, i.e. the worst case scenario for the user~\citep{toth_measuring_2004}.
The more general Rényi entropy has been proposed as a metric of anonymity
in~\foreignlanguage{british}{\citep{serjantov_towards_2003,bezzi_entropy_2008}}.
Neither of these have used entropy to measure privacy leakage in IDS
alarms.

\selectlanguage{british}%
The \paper{}  is also related to field of privacy-preserving intrusion
detection systems~\citep{SobireyHubnerPPIDS,HubnerIDA,SobireyAID,buschkes32,pang_high-level_2003,flegel_privacy-respecting_2007},
however neither of these solutions focus on privacy metrics.

\selectlanguage{english}%

\section{\label{sec:Conclusion}Conclusions}

In this \paper{}  we propose an entropy-based privacy leakage metric\index{privacy leakage metric}
founded on quantitative information flow\index{quantitative information flow}
analysis. An advantage is that this metric can be calculated based
on already existing information in the IDS alarm database. From a
privacy perspective, it provides a structured approach to identify
which IDS rules that may be leaking sensitive information and also
for handling such privacy leakages.

An advantage with the metric, is that it also is a measure of IDS
rule precision\index{IDS rule precision}. This is clearly desirable,
since the objective is to tune the IDS ruleset to reduce the leakage
of private or confidential information over time, for example through
improving the precision of the IDS rule or by applying anonymisation
techniques. This is also an advantage from a security perspective,
since more precise IDS rules mean less effort spent on false alarm
handling. 

We have demonstrated that the proposed approach is feasible based
on a set of real IDS alarms. It is furthermore shown that different
entropy algorithms and ways to calculate the standard deviation have
different strengths and weaknesses. Not surprisingly, the Laplacian
standard deviation based on the $L^{1}$ norm provides the most robust
statistic\index{robust statistic} to avoid problems with outliers\index{outliers},
a problem that has been shown to occur in the experimental data. The
experiments have shown that Shannon octet-entropy is the best entropy
metric with fastest convergence speed\index{convergence speed} for
reliably detecting encrypted traffic, and it is also the entropy metric
that is is best at distinguishing between plaintext and encrypted
traffic. It is also shown how the metric can avoid being incentive
incompatible by taking into account the length of the input data.

The Laplacian Mixture Model of the underlying data will in itself
be useful for classification\index{classification} purposes. If a
given model of the data has been identified, then this can be used
for subsequent classification of the underlying samples, for example
to anonymise IDS alarms from data clusters that may contain sensitive
information about user transactions, or to further classify the attack
vectors of the IDS alarms, for example to detect Denial of Service
attacks\index{Denial of Service attacks}. The clustering\index{clustering}
can therefore be used as a post-processing step to modify the IDS
alarms according to cluster, which means acting as a higher order
IDS solution.

A possible attack on the clustering method, is an overfitting attack\index{overfitting attack}
where a MSS provider decides to shirk\index{shirk} by deliberately
overfitting the attack vectors. The proposed method to avoid this,
is to ensure separation of duties\index{separation of duties} between
privacy and security interests and also that third party certification
organisations\index{certification organisations} oversee the operation.

The proposed privacy leakage metric only measures the primary privacy
leakage sources in IDS alarms. It does not consider secondary sources
of information leakage, like correlation of different information
sources. However, being able to measure the primary sources of privacy
leakage in IDS alarms is at least an initial approach that can and
should be considered before more elaborate analyses of the anonymity
set are performed. Furthermore, the ability to verify that the anonymisation
policies reduce measured information leakages means that policy verification\index{policy verification},
in the form of a privacy leakage gap analysis, will be possible in
order to provide incremental reductions of privacy leakage in IDS
alarms over time.

\section{\label{sec:Future-Work}Future Work}

Future work includes doing comparative studies of the performance
of different MSS providers from a privacy perspective. Adapting the
privacy leakage metric to support anomaly-based IDS is also left as
future work. This will amongst others require subdivision of the alarms,
for example based on service etc., to avoid that the entropy space
becomes too crowded by attack vectors.

Investigating possible secondary privacy leakages that may occur due
to inference or cross correlation between different information sources
both within the IDS alarm and outside is also left as future research.
This would require taking the privacy leakage metrics and evaluation
even further in order to evaluate the anonymity set that can be expected
for private or sensitive information, using metrics like differential
privacy~\citep{dwork_differential_2006,dwork_differential_2008,dwork_differential_2009},
k-anonymity \citep{de_capitani_di_vimercati_privacy_2006}, or l-diversity
\citep{MachanavajjhalaKiferGehrkeVenkitasubramaniam07_LDiversityPrivacyBeyoundKAnonymity}.

%\bibliographystyle{plainnat}
%\bibliography{bibliography,zotero,zotero-laptop}

\end{document}